\documentclass[12pt]{article}
\usepackage{amsfonts,graphicx,amssymb,amsmath,amsthm,epsfig}
\usepackage{float}
\allowdisplaybreaks

\begin{document}
\title{\bf Complexity of Charged Dynamical Spherical System in Modified Gravity}
\author{M. Sharif$^1$ \thanks{msharif.math@pu.edu.pk} and K. Hassan$^2$ \thanks{komalhassan3@gmail.com}\\
$^1$ Department of Mathematics and Statistics, The University of Lahore,\\
1-KM Defence Road Lahore, Pakistan.\\
$^2$ Department of Mathematics, University of the Punjab,\\
Quaid-e-Azam Campus, Lahore-54590, Pakistan.}

\date{}

\maketitle
\begin{abstract}
In this paper, we consider the effect of electromagnetic field to
the definition of complexity in the context of $f(G,T)$ gravity,
where $G$ and $T$ express the Gauss-Bonnet term and energy-momentum
tensor, respectively. The physical parameters such as anisotropic
pressure, charge, energy density inhomogeneity, heat dissipation and
correction terms are found responsible to induce complexity within
the self-gravitating objects. The scalar functions are determined
using Herrera's orthogonal splitting approach, which results in a
complexity factor that includes all of the system's essential
features. Furthermore, we investigate the dynamics of charged
spherical distribution by choosing homologous mode as the simplest
evolutionary pattern. Dissipative and non-dissipative cases
associated with complexity free and homologous conditions are also
discussed. Finally, we study the components responsible for
producing complexity during the evolution process. We deduce that
the inclusion of extra curvature terms and charge in $f(G, T)$
gravity enhances the complexity of the self-gravitating structure.
\end{abstract}

\section{Introduction}

The cosmos is made up of both microscopic and immense scale entities
that range from tiny planets to galaxies (comprising billions of
stars). These cosmic structures help in forming the basis for
cosmological studies and persuade the universe's progression.
Einstein believed that the cosmos is static but Hubble used a
relationship between the distance and recession velocity of galaxies
to prove that the universe is in an expanding phase. The rapid
expansion of the cosmos is believed to be caused by the mysterious
quantity termed as dark energy. Numerous astronomical phenomena like
supernovae and microwave background radiations \cite{1a,1b} indicate
the accelerated expanding universe. The $\Lambda$CDM model describes
the best paradigm for explaining the present accelerated expanding
universe but it faces two serious issues: cosmic coincidence and
fine-tuning. Consequently, modified theories of gravity are
introduced as alternative to discuss the cosmic rapid expansion.

Nojiri and Odintsov \cite{4} added the function of Gauss-Bonnet (GB)
invariant in the Einstein-Hilbert action and proposed modified GB
gravity, also known as $f(G)$ theory, where
$G=-4R^{\vartheta\tau}R_{\psi\chi}+R^{\vartheta\tau\alpha\delta}R_{\vartheta\tau\alpha\delta}+R^2$.
This theory discusses various astrophysical and cosmological events
happening in the universe. Bamba et al. \cite{5} chose a scale
factor as a conjunction of two exponential functions and
reconstructed the $f(G)$ model to analyze the early-time bounce and
late-time cosmic expansion. Abbas et al. \cite{5a} investigated the
possibility of constructing celestial objects and used the power-law
model together with Krori-Barua metric to inspect their physical
features in this theory. Sharif and Ikram \cite{5b} studied the
dynamics of inflationary phase of homogeneous and isotropic cosmos
utilizing viable $f(G)$ model and found various metric coefficients
that correspond to well-known inflationary models. Sharif and Ramzan
\cite{6b} used the embedding class-1 approach to develop the
anisotropic stellar structures and their salient physical aspects.

Sharif and Ikram \cite{7} formulated $f(G,T)$ theory by
incorporating trace of the energy-momentum tensor in $f(G)$ action.
They remodeled the power-law and de Sitter ansatz to discuss energy
conditions for the FRW cosmos. The same authors \cite{7a} discussed
the stable behavior of the Einstein universe through anisotropic
homogeneous perturbations. Hossienkhani et al. \cite{7b} analyzed
energy constraints of anisotropic universe and found a direct
relation between weak energy condition and anisotropy. Sharif and
Naeem \cite{8} studied the viable and stable behavior of certain
stellar quantities by formulating a new solution. Shamir \cite{8a}
determined the favorable bouncing solutions for the chosen equation
of state with respect to two $f(G,T)$ gravity models.

The electromagnetic field plays an important role (due to the
suppression of gravitational force) in exploring the stability and
evolution of massive bodies. To maintain the stable character of
astrophysical objects, a substantial quantity of charge is essential
to overcome the gravitational pull. Ivanov \cite{9a} investigated
three distinct classes of regular and general solutions to the field
equations for the charged celestial body. Esculpi and Aloma
\cite{9ab} used a linear equation of state that connects tangential
and radial pressures to study the impact of anisotropic factor on
charged compact objects. Sharif and Naz \cite{9b} inspected the
collapse of a cylindrical charged configuration using dynamical
field equations in $f(G)$ gravity. We have figured out the essential
characteristics of uncharged/charged and charged anisotropic
celestial objects through minimal and extended geometric deformation
scheme, respectively, in $f(G,T)$ gravity \cite{9c,9d,9f}.

The complexity of astrophysical entities predominantly relies on the
internal attributes ( heat, pressure, temperature, energy density,
etc.) of self-gravitating objects. In order to calculate the
complexity of cosmic objects, a mathematical formula that
incorporates all important physical variables must be applied.
L{\'o}pez-Ruiz and his collaborators \cite{10a} proposed the idea of
complexity in terms of information and entropy. The ideal gas and
perfect crystal were the first two physical quantities on which this
notion was used. The perfect crystal has no entropy due to the
symmetry of molecules while the ideal gas exhibits maximal entropy
due to the random distribution of its molecules. The probability
distribution of perfect crystal does not reveal much new
information, thus a tiny segment describes all of its important
features. In contrast to perfect crystal, we attain a large amount
of information during the study of small portion of ideal gas. One
can notice that both the entities are at their extremes in behavior,
so zero complexity is allotted to them. A new concept of complexity
was developed based on how various probabilistic states deviate from
the equiprobable distribution of the system \cite{10b,10c}.
According to this notion, perfect crystal and ideal gas also have
zero complexity.

Herrera \cite{12} revised the concept of complexity in such a way
that it incorporates all the essential state determinants (pressure
anisotropy, inhomogeneous energy density, Tolman mass) for the
static symmetric anisotropic matter distribution. This technique
works on splitting the Riemann tensor which gives rise to the
formulation of various scalar functions. Thus, we call a factor as
complexity producing source that includes all the above-mentioned
quantities. Sharif and Butt \cite{16} discussed the complexity
factor for the charged sphere. This notion of complexity was then
extended to non-static dissipative fluid by Herrera et al \cite{13}
and discussed the simplest evolutionary mode. The application of
this concept has also been found in axially symmetric spacetime by
the same authors \cite{14}. Sharif and Butt \cite{15} employed the
Herrera's strategy to gauge the complexity of static cylindrically
anisotropic source.

The analysis of complexity has also been investigated in modified
theories. Sharif et al. \cite{17a} discussed complexity through
${\mathcal{Y}}_{TF}$ (the complexity factor) for spherical system in
$f(G)$ gravity. Yousaf et al. \cite{17b,17ab} studied the the
complexity producing factor for the charged and uncharged spherical
geometry in $f(G,T)$ gravity. We have also figured out the impact of
complexity on static uncharged/charged cylindrical source and
non-static (spherical and cylindrical) structures in $f(G,T)$
gravity through the orthogonal splitting of the Riemann tensor
\cite{17g,19,19a,19ab}. In the background of different modified
theories, there is a substantial body of work on the assessment of
complexity in diverse geometries \cite{18,26,27,28,29,31,32}.

This paper explores the complexity producing factor for the
dynamical charged spherical configuration as well as evolutionary
patterns in $f(G,T)$ theory. The paper is arranged in the following
format. The fundamental properties of the matter source and the
field equations characterizing the evolution of the system are given
in section \textbf{2}. The Riemann tensor is orthogonally decomposed
into its structure scalars in section \textbf{3}. In section
\textbf{4}, two patterns of evolution, i.e., homologous and
homogeneous are investigated. In section \textbf{5}, we derive
kinematical and dynamical quantities to examine possible solutions
for dissipative and non-dissipative systems. The conditions under
which self-gravitating objects deviate from zero complexity
condition are observed in section \textbf{6}. The summary of the
obtained results is presented in section \textbf{7}.

\section{$f(G,T)$ Gravity and Matter Determinants}

The generic $f(G,T)$ function along with Ricci scalar in
Einstein-Maxwell action takes the form
\begin{equation}\label{1}
\mathbf{S}_{f(G,T)}=\frac{1}{2k^2}\int d^{4}x[f(G,T)+R]\sqrt{-g}
+\int\sqrt{-g}(\mathbb{L}_m+\mathbb{L}_E)d^{4}x,
\end{equation}
where the Lagrangian densities for the normal matter and
electromagnetic field are represented by $\mathbb{L}_m$ and
$\mathbb{L}_E$, respectively, and $k^{2}$ denotes the coupling
constant. The stress-energy tensor and Lagrangian density are
interlinked through the relation
\begin{equation}\label{1a}
T_{\vartheta\tau}=g_{\vartheta\tau}\mathbb{L}_m-\frac{2\partial\mathbb{L}_m}{\partial
g^{\vartheta\tau}}.
\end{equation}
Here, the variation of action \eqref{1} with the metric tensor
$(g_{\vartheta\tau})$ formulates the modified field equations as
\begin{eqnarray}\nonumber
G_{\vartheta\tau}&=&8\pi
(T_{\vartheta\tau}+\verb"S"_{\vartheta\tau})-(\Theta_{\vartheta\tau}+T_{\vartheta\tau})f_{T}(G,T)
+\frac{1}{2}g_{\vartheta\tau}f(G,T)+(4R_{\mu\tau}R^{\mu}_{\vartheta}\\\nonumber
&+&4R^{\mu\nu}R_{\vartheta\mu \tau \nu}-2RR_{\vartheta\tau}-2R^{\mu
\nu \gamma} _{\vartheta}R_{\tau \mu \nu
\gamma})f_{G}(G,T)+(4g_{\vartheta\tau}R^{\mu
\nu}\nabla_{\mu}\nabla_{\nu}\\\nonumber
&+&2R\nabla_{\vartheta}\nabla_{\tau}+4R_{\vartheta\tau}\nabla^{2}-2g_{\vartheta\tau}R\nabla^{2}
-4R^{\mu}_{\vartheta}\nabla_{\tau}\nabla_{\mu}
-4R^{\mu}_{\tau}\nabla_{\vartheta}\nabla_{\mu}\\\label{2}&-&4R_{\vartheta\mu
\tau \nu}\nabla^{\mu}\nabla^{\nu}) f_{G}(G,T),
\end{eqnarray}
where d' Alembert operator is described by
$\nabla^{2}=\nabla^{l}\nabla_{l}=\Box$ and
$\Theta_{\vartheta\tau}=-2T_{\vartheta\tau}+\mathfrak{p}g_{\vartheta\tau}$.
The Einstein tensor is indicated by
$G_{\vartheta\tau}=R_{\vartheta\tau}-\frac{1}{2}Rg_{\vartheta\tau}$,
$f_{G}(G,T)$ and $f_{T}(G,T)$ are the partial derivatives of
$f(G,T)$ (arbitrary function) with respect to $G$ and $T$,
respectively.

In this theory, the matter components are coupled with the
geometrical terms that lead to the non-conservation of the
stress-energy tensor. Due to this, an extra force is induced that
forces the particles in the gravitational field to deviate from the
geodesic trajectory, therefore, the covariant differentiation of
Eq.\eqref{2} reads
\begin{eqnarray}\nonumber
\nabla^{\vartheta}T_{\vartheta\tau}&=&\frac{f_{T}(G,T)}{k^{2}-f_{T}(G,T)}
\left[-\frac{1}{2}g_{\vartheta\tau}\nabla^{\vartheta}T+(\Theta_{\vartheta\tau}+T_{\vartheta\tau})\nabla^{\vartheta}(\ln
f_{T}(G,T))\right.\\\label{2a}&+&
\left.\nabla^{\vartheta}\Theta_{\vartheta\tau}\right].
\end{eqnarray}
Equation \eqref{2} can also be written as
\begin{equation}\label{3a}
G_{\vartheta\tau}=8\pi
T^{\textsf{(tot)}}_{\vartheta\tau}=8\pi(T^{(M)}_{\vartheta\tau}+\verb"S"_{\vartheta\tau}+T^{\textsf{(GT)}}_{\vartheta\tau}),
\end{equation}
where $T^{\textsf{(GT)}}_{\vartheta\tau})$ are extra curvature terms
of $f(G,T)$ given as
\begin{eqnarray}\nonumber
T^{\textsf{(GT)}}_{\vartheta\tau}&=&\frac{1}{8\pi}\left[\{(\mathfrak{u}+\mathfrak{p})v_{\vartheta}v_{\tau}
+\Pi_{\vartheta\tau}+\zeta(v_{\vartheta}\chi_{\tau}+\chi_{\vartheta}v_{\tau})\}f_{T}(G,T)
\right.\\\nonumber
&+&\left.(4R_{\mu\tau}R^{\mu}_{\vartheta}+4R^{\mu\nu}R_{\vartheta\mu
\tau \nu}-2RR_{\vartheta\tau}-2R^{\mu \nu \gamma}
_{\vartheta}R_{\tau \mu \nu \gamma})f_{G}(G,T)\right.\\\nonumber
&+&\left.(4g_{\vartheta\tau}R^{\mu
\nu}\nabla_{\nu}\nabla_{\mu}-4R_{\vartheta\mu
\tau\nu}\nabla^{\nu}\nabla^{\mu}-4R^{\mu}_{\vartheta}\nabla_{\tau}\nabla_{\mu}-2g_{\vartheta\tau}R\nabla^{2}\right.\\\label{4a}
&-&\left.4R^{\mu}_{\tau}\nabla_{\vartheta}\nabla_{\mu}+2R\nabla_{\vartheta}\nabla_{\tau}
+4R_{\vartheta\tau}\nabla^{2})f_{G}(G,T)\right]+\frac{g_{\vartheta\tau}f(G,T)}{2}.
\end{eqnarray}
The normal matter distribution possessing anisotropy and dissipation
due to heat flux is followed by the energy-momentum tensor
\begin{equation}\label{5a}
T^{(M)}_{\vartheta\tau} =\mathfrak{u}
v_{\vartheta}v_{\tau}+{\mathfrak{p}}h_{\vartheta\tau}
+\Pi_{\vartheta\tau}+{\zeta}(v_{\vartheta}\chi_{\tau}+\chi_{\vartheta}v_{\tau}),
\end{equation}
where $h_{\vartheta\tau}$ denotes the projection tensor. The radial
four-vector, four velocity and heat flux are expressed as
$\chi^{\vartheta}=\left(0,\textsf{B}^{-1},0,0\right)$,~$v^{\vartheta}=\left(\textsf{A}^{-1},0,0,0\right)$
and ${\zeta}^{\vartheta}=\left(0,{\zeta}\textsf{B}^{-1},0,0\right)$,
respectively. These quantities satisfy the following relations
\begin{eqnarray}\label{101c}
\chi^{\vartheta}\chi_{\vartheta}&=&1,\quad
\chi^{\vartheta}v_{\vartheta}=0,\quad
v^{\vartheta}{\zeta}_{\vartheta}=0, \quad
v^{\vartheta}v_{\vartheta}=-1.
\end{eqnarray}
The remaining entities are delineated as
\begin{eqnarray}\label{101a}
\Pi_{\vartheta\tau}&=&\Pi\left(\chi_{\vartheta}\chi_{\tau}-\frac{h_{\vartheta\tau}}{3}\right),
\quad  \Pi={\mathfrak{p}}_{r}-{\mathfrak{p}}_{\bot},\\\label{101b}
{\mathfrak{p}}&=&\frac{{\mathfrak{p}}_{r}+2{\mathfrak{p}}_{\bot}}{3},
\quad h_{\vartheta\tau}=g_{\vartheta\tau}+v_{\vartheta}v_{\tau}.
\end{eqnarray}
The stress-energy tensor in the electromagnetic field reads
\begin{equation}\label{3b}
\verb"S"_{\vartheta\tau}=\frac{1}{4\pi}\left(\mathcal{F}^{l}_{\vartheta}\mathcal{F}_{\tau
l}-\frac{1}{4}g_{\vartheta\tau}\mathcal{F}_{lm}\mathcal{F}^{lm}\right),
\end{equation}
where $\mathcal{F}_{\vartheta\tau}$ signifies the Maxwell field
tensor. In tensorial form, the Maxwell field equations are described
as
\begin{equation}\nonumber
\mathcal{F}_{[\vartheta\tau;l]}=0,\quad
\mathcal{F}^{\vartheta\tau}_{~~;\tau}=4\pi \mathcal{J}^{\vartheta},
\end{equation}
where $\mathcal{J}^{\vartheta}=\varepsilon v^{\vartheta}$ is the
four current and $\varepsilon$ indicates the charge density.

The line element representing the non-static geometrical structure
(surrounded by a hypersurface $\Sigma$) involving dissipation is
provided by
\begin{equation}\label{8a}
ds^{2}=-\textsf{A}^{2}dt^{2}+{\textsf{B}}^{2}dr^{2}+{\textsf{C}}^{2}d\theta^{2}+{\textsf{C}}^2{\sin^{2}\theta}{d\phi^2},
\end{equation}
where {\textsf{B}}={\textsf{B}}(t,r)$,~
${\textsf{A}}={\textsf{A}}(t,r) and
${\textsf{C}}={\textsf{C}}(t,r)$. The modified field equations for
the non-static realm are computed as
\begin{eqnarray}\label{9a}
8\pi({\textsf{A}}^{2}{\mathfrak{u}}+\frac{\mathbf{s}^2{\textsf{A}}}{8\pi
\textsf{C}^4}+T^{\textsf{(GT)}}_{00}) &=&
\frac{-{\textsf{A}}^2\left[\frac{2{\textsf{C}}''}{{\textsf{C}}}+\frac{{\textsf{C}}'^2}{{\textsf{C}}^2}-\frac{{\textsf{B}}^2}{{\textsf{C}}^2}
-\frac{2{\textsf{C}}'{\textsf{B}}'}{{\textsf{C}}{\textsf{B}}}\right]}{{\textsf{B}}^2}
+\frac{\dot{{\textsf{C}}}(\frac{2\dot{{\textsf{B}}}}{{\textsf{B}}}+\frac{\dot{{\textsf{C}}}}{{\textsf{C}}})}{{\textsf{C}}},\\\label{10a}
8\pi (-{\zeta}{\textsf{A}}{\textsf{B}}+T^{\textsf{(GT)}}_{01}) &=&
\frac{2{\textsf{A}}'\dot
{\textsf{C}}}{{\textsf{A}}{\textsf{C}}}+\frac{2{\textsf{C}}'\dot
{\textsf{B}}}{{\textsf{C}}{\textsf{B}}}-\frac{2\dot
{\textsf{C}}'}{{\textsf{C}}},
\\\nonumber 8\pi({\textsf{B}}^{2}{\mathfrak{p}}_{r}-\frac{\mathbf{s}^2{\textsf{B}}^{2}}{8\pi
\textsf{C}^4}+T^{\textsf{(GT)}}_{11})&=&
\frac{-{\textsf{B}}^2\left[-\frac{\dot{{\textsf{C}}}}{{\textsf{C}}}(\frac{2\dot{{\textsf{A}}}}{{\textsf{A}}}
-\frac{\dot{{\textsf{C}}}}{{\textsf{C}}})+\frac{2\ddot{{\textsf{C}}}}{{\textsf{C}}}\right]}{{\textsf{A}}^2}
-\frac{{\textsf{B}}^2}{{\textsf{C}}^2}+(\frac{2{\textsf{A}}'{\textsf{C}}'}{{\textsf{A}}{\textsf{C}}}+\frac{{\textsf{C}}'^2}{{\textsf{C}}^2}
),\\\label{11a}\\\nonumber
8\pi({\textsf{C}}^{2}{\mathfrak{p}}_\bot+\frac{\mathbf{s}^2}{8\pi
\textsf{C}^2}+T^{\textsf{(GT)}}_{22})&=&
\frac{-{\textsf{C}}^2}{{\textsf{A}}^2}\left[\frac{\dot{{\textsf{B}}}\dot{{\textsf{C}}}}{{\textsf{B}}{\textsf{C}}}
+\frac{\ddot {\textsf{B}}}{{\textsf{B}}}+\frac{\ddot
{\textsf{C}}}{{\textsf{C}}}-\frac{\dot
{\textsf{A}}}{{\textsf{A}}}(\frac{\dot
{\textsf{C}}}{{\textsf{C}}}+\frac{\dot
{\textsf{B}}}{{\textsf{B}}})\right]\\\label{12a}
&+&\frac{{\textsf{C}}^2}{{\textsf{B}}^2}\left[(\frac{{\textsf{A}}'}{{\textsf{A}}}-\frac{{\textsf{B}}'}{{\textsf{B}}})\frac{{\textsf{C}}'}{{\textsf{C}}}
-\frac{{\textsf{A}}'{\textsf{B}}'}{{\textsf{A}}
{\textsf{B}}}+\frac{{\textsf{A}}''}{{\textsf{A}}}+\frac{{\textsf{C}}''}{{\textsf{C}}}\right],
\end{eqnarray}
where temporal and radial partial derivatives are denoted by . and
$\prime$, respectively. The expressions for modified terms
($T^{\textsf{(GT)}}_{00},~T^{\textsf{(GT)}}_{01},~T^{\textsf{(GT)}}_{11}~
T^{\textsf{(GT)}}_{22}$) are presented in
Eqs.(\ref{100})-(\ref{100c}) of Appendix \textbf{A}. Using
Eq.\eqref{2a}, the non-zero components of the Bianchi identities for
the present setup are derived as
\begin{align}\label{13a}
T^{\vartheta\tau}_{;\tau}v_{\vartheta}&=\frac{-1}{{\textsf{A}}}\bigg\{\dot{\mathfrak{u}}
+\frac{\dot
{\textsf{B}}}{{\textsf{B}}}\left({\mathfrak{u}}+{\mathfrak{p}}_{r}\right)+2\frac{\dot
{\textsf{C}}}{{\textsf{C}}}\left({\mathfrak{u}}+{\mathfrak{p}}_{\bot}\right)\bigg\}
-\frac{1}{{\textsf{B}}}\bigg\{{\zeta}'+2{\zeta}\bigg(\frac{\textsf{A}'}{\textsf{A}}+\frac{\textsf{C}'}{\textsf{C}}\bigg)
\bigg\}=\texttt{Z}_{1},
\\\nonumber
T^{\vartheta\tau}_{;\tau}\chi_{\vartheta}&=\frac{1}{{\textsf{A}}}\bigg\{\dot
{\zeta} +2{\zeta}(\frac{\dot {\textsf{B}}}{{\textsf{B}}}+\frac{\dot
{\textsf{C}}}{{\textsf{C}}})\bigg\}+\frac{1}{{\textsf{B}}}\bigg\{{\mathfrak{p}}'_{r}
+\left({\mathfrak{u}}+{\mathfrak{p}}_{r}
\right)\frac{{\textsf{A}}'}{{\textsf{A}}}-\frac{\mathbf{s}\mathbf{s}'}{4\pi
\textsf{C}^4}+\left(2{\mathfrak{p}}_{r}-2{\mathfrak{p}}_{\bot}
\right)\\\label{14a}&\times\frac{{\textsf{C}}'}{{\textsf{C}}}\bigg\}=\texttt{Z}_{2},
\end{align}
where the extra curvature terms $\texttt{Z}_{1}$ and
$\texttt{Z}_{2}$ are prescribed in Eqs.(\ref{100d}) and (\ref{100e})
of Appendix \textbf{A}.

The non-zero elements of shear tensor, four acceleration and
expansion scalar, respectively, for the present distribution are
defined as
\begin{equation}\label{18a}
\sigma_{11}=\frac{2}{3}{\textsf{B}}^2\sigma, \quad
\sigma_{22}=\frac{\sigma_{33}}{\sin^2
\theta}=-\frac{1}{3}{\textsf{C}}^2\sigma,
\end{equation}
\begin{equation}\label{19a}
\sigma^{\vartheta\tau}\sigma_{\vartheta\tau}=\frac{2}{3}\sigma^2,~\sigma=\left(\frac{\dot
{\textsf{B}}}{{\textsf{B}}}-\frac{\dot
{\textsf{C}}}{{\textsf{C}}}\right)\frac{1}{{\textsf{A}}},
\end{equation}
\begin{equation}\label{15a}
a=\sqrt{a^{\vartheta}a_{\vartheta}}=\frac{{\textsf{A}}'}{{\textsf{B}}{\textsf{A}}},
\quad a_{1}=\frac{{\textsf{A}}'}{{\textsf{A}}},
\end{equation}
\begin{equation}\label{17a}
\Theta=\left(2\frac{\dot {\textsf{C}}}{{\textsf{C}}}+\frac{\dot
{\textsf{B}}}{{\textsf{B}}}\right)\frac{1}{{\textsf{A}}}.
\end{equation}
The influence of shear and expansion scalars on the fluid matter
source is investigated through Eq.\eqref{10a} in the following form
\begin{equation}\label{21a}
4\pi\left({\zeta}{\textsf{B}}-\frac{T^{\textsf{(GT)}}_{01}}{{\textsf{A}}}\right)=-\sigma
\frac{{\textsf{C}}'}{{\textsf{C}}}+\frac{1}{3}\left(\Theta-\sigma\right)'=\frac{{\textsf{C}}'}{{\textsf{B}}}
\left[\frac{1}{3}D_{{\textsf{C}}}\left(\Theta-\sigma\right)
-\frac{\sigma}{{\textsf{C}}}\right],
\end{equation}
where proper radial differentiation is indicated by
$D_{{\textsf{C}}}=\frac{1}{{\textsf{C}}'}\frac{\partial}{\partial
r}$. The mass of inner spherical object can be employed by Misner
and Sharp \cite{19b} defined as
\begin{equation}\label{22a}
m=\frac{1}{2}{\textsf{C}}^3R_{232}^{3}=\left[1-\left(\frac{{\textsf{C}}'}{{\textsf{B}}}\right)^2+\left(\frac{\dot
{\textsf{C}}}{{\textsf{A}}}\right)^2\right]\frac{{\textsf{C}}}{2}+\frac{\mathbf{s}^2}{2\textsf{C}}.
\end{equation}
Here, the dynamics of self-gravitating structure is examined by
introducing proper time derivative , i.e.,
$D_{T}=\frac{1}{{\textsf{A}}}\frac{\partial}{\partial t}$. During
collapse, the outward pressure is suppressed by gravity which
results in the decrement of radius of any celestial object. Due to
this, the velocity $(U)$ of the collapsing fluid in terms of areal
radius (${\textsf{C}}$) becomes negative as
\begin{equation}\label{23a}
U=D_{T}{\textsf{C}}< 0.
\end{equation}
The mass and velocity of the charged matter source are interlinked
as follows
\begin{equation}\label{25a}
E\equiv\frac{{\textsf{C}}'}{{\textsf{B}}}=\left(1-\frac{2m}{{\textsf{C}}}+\frac{\mathbf{s}^2}{2\textsf{C}}+U^2\right)^\frac{1}{2}.
\end{equation}
In view of time and radial proper derivatives, the mass function
takes the form
\begin{equation}\label{27a}
D_{T}m=-4\pi\left[\left({\mathfrak{p}}_{r}-\frac{\mathbf{s}^2}{8\pi\textsf{C}^4}+\frac{T^{\textsf{(GT)}}_{11}}{{\textsf{B}}^2}\right)U
+\left({\zeta}-\frac{T^{\textsf{(GT)}}_{01}}{{\textsf{A}}{\textsf{B}}}\right)E\right]{\textsf{C}}^2+\frac{\mathbf{s}\dot{\mathbf{s}}}{\textsf{A}\textsf{C}}
-\frac{\mathbf{s}^2\dot{\textsf{C}}}{2\textsf{A}\textsf{C}^2},
\end{equation}
\begin{equation}\label{28a}
D_{{\textsf{C}}}m=4\pi\left[\left({\mathfrak{u}}+\frac{\mathbf{s}^2}{8\pi\textsf{C}^4}
+\frac{T^{\textsf{(GT)}}_{00}}{{\textsf{A}}^2}\right)
+\left({\zeta}-\frac{T^{\textsf{(GT)}}_{01}}{{\textsf{A}}{\textsf{B}}}\right)\frac{U}{E}\right]
{\textsf{C}}^2+\frac{\mathbf{s}\mathbf{s}'}{\textsf{C}\textsf{C}'}
-\frac{\mathbf{s}^2}{2\textsf{C}^2},
\end{equation}
which yield
\begin{eqnarray}\nonumber
\frac{3m}{{\textsf{C}}^3}&=&4\pi\left({\mathfrak{u}}+\frac{\mathbf{s}^2}{8\pi\textsf{C}^4}+\frac{T^{\textsf{(GT)}}_{00}}{{\textsf{A}}^2}\right)
-\frac{4\pi}{{\textsf{C}}^3}\int^{r}_{0}{\textsf{C}}^3\left[D_{{\textsf{C}}}\left({\mathfrak{u}}+\frac{\mathbf{s}^2}{8\pi\textsf{C}^4}
+\frac{T^{\textsf{(GT)}}_{00}}{{\textsf{A}}^2}\right)
\right.\\\label{30a}&-&\left.3\left({\zeta}-\frac{T^{\textsf{(GT)}}_{01}}{{\textsf{A}}{\textsf{B}}}\right)
\frac{U}{{\textsf{C}}E}\right]{\textsf{C}}'dr+\frac{3\mathbf{s}^2}{2\textsf{C}^4}.
\end{eqnarray}

The deformation induced in a self-gravitating body (as a result of
varying gravitational field) of nearby object is estimated by the
Weyl tensor
$({\textsf{C}}^{\lambda}_{\vartheta\tau{\mathfrak{u}}})$. This
tensor is split into two components, i.e., electric and magnetic.
For the spherical geometry, the magnetic component appears to be
zero while electric part takes the following form
\begin{equation}\label{35a}
E_{\vartheta\tau}={\textsf{C}}_{\vartheta{\mathfrak{u}}\tau\nu}v^{{\mathfrak{u}}}v^{\nu}
=\epsilon(-\frac{h_{\vartheta\tau}}{3}+\chi_{\vartheta}\chi_{\tau}),
\end{equation}
where the Weyl scalar reads
\begin{eqnarray}\nonumber
\epsilon &=&\frac{1}{2{\textsf{A}}^2}\left[\frac{\ddot
{\textsf{C}}}{{\textsf{C}}}-\frac{\ddot
{\textsf{B}}}{{\textsf{B}}}-\left(\frac{\dot
{\textsf{C}}}{{\textsf{C}}}+\frac{\dot
{\textsf{A}}}{{\textsf{A}}}\right)\left(\frac{\dot
{\textsf{C}}}{{\textsf{C}}}-\frac{\dot
{\textsf{B}}}{{\textsf{B}}}\right)\right]-\frac{1}{2{\textsf{C}}^2}+\frac{1}{2{\textsf{B}}^2}\left[-\frac{{\textsf{C}}''}{{\textsf{C}}}+\frac{{\textsf{A}}''}{{\textsf{A}}}
\right.\\\label{36a}
&+&\left.\left(\frac{{\textsf{C}}'}{{\textsf{C}}}
-\frac{{\textsf{A}}'}{{\textsf{A}}}\right)\left(\frac{{\textsf{B}}'}{{\textsf{B}}}+\frac{{\textsf{C}}'}{{\textsf{C}}}\right)\right].
\end{eqnarray}
The following relation explains the influence of tidal force on the
charged matter distribution as
\begin{equation}\label{46a}
\frac{3m}{{\textsf{C}}^3}=4\pi\left[-\Pi^{\textsf{(tot)}}+\left({\mathfrak{u}}+\frac{3\mathbf{s}^2}{8\pi\textsf{C}^4}
+\frac{T^{\textsf{(GT)}}_{00}}{{\textsf{A}}^2}\right)
\right]+\frac{3\mathbf{s}^2}{2\textsf{C}^4}-\epsilon,
\end{equation}
where $\Pi^{\textsf{(tot)}}=\Pi^{\textsf{(GT)}}+\Pi$.

\section{Structure Scalars}

The concept of orthogonal splitting of the Riemann tenor was
introduced by Bel \cite{22b}. Following this procedure, Herrera
\cite{24a} developed different scalar functions to devise complexity
of the system named as structure scalars. The Ricci tensor, Weyl
tensor and Ricci scalar are defined in terms of the Riemann tensor
as
\begin{equation}\label{90a}
R^{\rho}_{\vartheta\tau\mu}={\textsf{C}}^{\rho}_{\vartheta\tau\mu}
+\frac{1}{2}R_{\vartheta\mu}\delta^{\rho}_{\tau}
+\frac{1}{2}R^{\rho}_{\tau}g_{\vartheta\mu}
-\frac{1}{2}R_{\vartheta\tau}\delta^{\rho}_{\mu}-\frac{1}{2}R^{\rho}_{\mu}g_{\vartheta\tau}
-\frac{1}{6}R\left(\delta^{\rho}_{\tau}g_{\vartheta\mu}\right.
-\left.g_{\vartheta\tau}\delta^{\rho}_{\mu}\right),
\end{equation}
which can also be expressed in the form of matter source as
\begin{equation}\label{39a}
R^{\vartheta\gamma}_{\tau\delta}={\textsf{C}}^{\vartheta\gamma}_{\tau\delta}+16\pi
T^{\textsf{(tot)}[\vartheta}_{[\tau}\delta^{\gamma]}_{\delta]}+8\pi
T^{\textsf{(tot)}}\left(\frac{1}{3}\delta^{\vartheta}_{[\tau}\delta^{\gamma}_{\delta]}-\delta^{[\vartheta}_{[\tau}\delta^{\gamma]}_{\delta]}\right).
\end{equation}
Here, we propose two tensors $X_{\vartheta\tau}$ and
$Y_{\vartheta\tau}$ defined as
\begin{eqnarray}\label{37a}
X_{\vartheta\tau}
&=&^{\ast}R^{\ast}_{\vartheta\gamma\tau\delta}v^{\gamma}v^{\delta}=\frac{1}{2}\eta^{\epsilon\mu}_{\vartheta\gamma}
R^{\ast}_{\epsilon\mu\tau\delta}v^{\gamma}v^{\delta},
\\\label{38a}Y_{\vartheta\tau}
&=&R_{\vartheta\gamma\tau\delta}v^{\gamma}v^{\delta},
\end{eqnarray}
where $R^{\ast}_{\vartheta\tau\gamma\delta}
=\frac{1}{2}\eta_{\epsilon\mu\gamma\delta}R^{\epsilon\mu}_{\vartheta\tau}$
and $\eta^{\epsilon\mu}_{\vartheta\gamma}$ indicates the Levi-Civita
symbol. One can express these tensors into four scalars, i.e., a
combination of trace $({\mathcal{Y}}_{T}, {\mathcal{X}}_{T})$ and
trace-free parts $({\mathcal{Y}}_{TF}, {\mathcal{X}}_{TF})$ as
\begin{eqnarray}\label{40a}
Y_{\vartheta\tau}
&=&(\chi_{\vartheta}\chi_{\tau}-\frac{h_{\vartheta\tau}}{3}){\mathcal{Y}}_{TF}+\frac{h_{\vartheta\tau}{\mathcal{Y}}_{T}}{3},
\\\label{41a}
X_{\vartheta\tau}
&=&(\chi_{\vartheta}\chi_{\tau}-\frac{h_{\vartheta\tau}}{3}){\mathcal{X}}_{TF}+\frac{h_{\vartheta\tau}{\mathcal{X}}_{T}}{3}.
\end{eqnarray}

For the considered setup, the four structure scalars turn out to be
\begin{eqnarray}\label{42a}
{\mathcal{Y}}_{T}&=&
\frac{\mathbf{s}^2}{\textsf{C}^4}+4\pi\left({\mathfrak{u}}+3{\mathfrak{p}}_{r}-2\Pi
\right)+\frac{\left({\mathfrak{u}}+{\mathfrak{p}}\right)f_{T}}{2}+\textsf{M}^{\textsf{(GT)}},
\\\label{43a} {\mathcal{Y}}_{TF} &=& \epsilon-4\pi
\Pi-\frac{\mathbf{s}^2}{\textsf{C}^4}+\frac{\Pi}{2}f_{T}+\textsf{L}^{\textsf{(GT)}},\\\label{44a}
{\mathcal{X}}_{T}&=& 8\pi
({\mathfrak{u}}+\frac{\mathbf{s}^2}{8\pi\textsf{C}^4})+\textsf{Q}^{\textsf{(GT)}},
\\\label{45a} {\mathcal{X}}_{TF} &=&-\epsilon-4\pi \Pi-\frac{\mathbf{s}^2}{\textsf{C}^4}+\frac{\Pi}{2}f_{T},
\end{eqnarray}
where
$\textsf{L}^{\textsf{(GT)}}=\frac{\textsf{J}^{\textsf{(GT)}}_{\vartheta\tau}}{\chi_{\vartheta}\chi_{\tau}-\frac{1}{3}h_{\vartheta\tau}}$.
The extra terms in scalar functions ($\textsf{M}^{\textsf{(GT)}}$,
$\textsf{J}^{\textsf{(GT)}}_{\vartheta\tau}$,\\$\textsf{Q}^{\textsf{(GT)}}$)
are given in Appendix \textbf{B}. The contribution of energy density
and charge in the system is attributed to the scalar
${\mathcal{X}}_{T}$ whereas the anisotropic stresses along with
electromagnetic effects are controlled by ${\mathcal{Y}}_{T}$ . The
scalar function ${\mathcal{Y}}_{TF}$ in terms of all physical
parameters is obtained using Eqs.\eqref{36a} and \eqref{43a} as
\begin{eqnarray}\nonumber
{\mathcal{Y}}_{TF}&=&\frac{4\pi}{{\textsf{C}}^3} \int
{\textsf{C}}^3\left[D_{{\textsf{C}}}\left({\mathfrak{u}}+\frac{\mathbf{s}^2}{8\pi\textsf{C}^4}+\frac{T^{\textsf{(GT)}}_{00}}{{\textsf{A}}^2}\right)
-3\left({\zeta}-\frac{T^{\textsf{(GT)}}_{01}}{{\textsf{A}}{\textsf{B}}}\right)\frac{U}{{\textsf{C}}E}\right]{\textsf{C}}'dr
\\\label{47a}&-&4\pi\Pi-4\pi\Pi^{\textsf{(GT)}}-\textsf{L}^{\textsf{(GT)}}-\frac{\mathbf{s}^2}{2\textsf{C}^4}+\frac{\Pi}{2}f_{T}.
\end{eqnarray}
One can notice from the above equation that the scalar
${\mathcal{Y}}_{TF}$ is composed of heat flux, inhomogeneous energy
density, modified terms, charge terms and pressure anisotropy . On
the other hand, ${\mathcal{X}}_{TF}$ includes the involvement of
inhomogeneity due to energy density together with additional terms
and charge as
\begin{equation}\label{48a}
{\mathcal{X}}_{TF}=4\pi\Pi^{\textsf{(GT)}}-\frac{2\mathbf{s}^2}{\textsf{C}^4}-\frac{4\pi}{{\textsf{C}}^3}\int
{\textsf{C}}^3\bigg[D_{{\textsf{C}}}\bigg({\mathfrak{u}}+\frac{\mathbf{s}^2}{8\pi\textsf{C}^4}+\frac{T^{\textsf{(GT)}}_{00}}{{\textsf{A}}^2}\bigg)
-3\bigg({\zeta}-\frac{T^{\textsf{(GT)}}_{01}}{{\textsf{A}}{\textsf{B}}}\bigg)
\frac{U}{{\textsf{C}}E}\bigg]{\textsf{C}}'dr.
\end{equation}

\section{Evolution Modes}

The physical parameters (pressure, charge and energy density) have a
significant impact on how the cosmic structures behave. The scalar
${\mathcal{Y}}_{TF}$ involves energy density inhomogeneity, heat
flux and pressure anisotropy along with $f(G,T)$ corrections. The
complexity factor for the anisotropic dissipative massive source in
a non-static geometry is thus identified as ${\mathcal{Y}}_{TF}$. To
obtain the complexity-free structure, we substitute
${\mathcal{Y}}_{TF}=0$ which is regarded as complexity-free
condition. As the fluid under consideration is evolving with time,
thus it is necessary to investigate the patterns of evolution. In
the subsequent sections, two evolutionary modes, i.e., homologous
evolution as well as homogeneous expansion are examined. We also
determine the simplest mode that can reduce complexity in the
evolutionary process.

\subsection{Homologous Evolution}

The homologous refers to the phenomenon in which the whole system
depicts a similar pattern. The collapse happens when all the
substance of any astronomical object descends towards the center in
the interior region. A direct relationship between radial distance
and velocity is observed in homologous collapse, implying that
during the collapse, all material of the astrophysical entity falls
into its core at the same rate. In contrast to homologously
collapsing objects, those structures whose cores collapse first will
thus generate less gravitational radiations. Equation \eqref{21a}
can also be written as
\begin{equation}\label{52a}
D_{{\textsf{C}}}\left(\frac{U}{{\textsf{C}}}\right)=\frac{\sigma}{{\textsf{C}}}+\frac{4\pi}{E}\left({\zeta}
-\frac{T^{\textsf{(GT)}}_{01}}{{\textsf{A}}{\textsf{B}}}\right),
\end{equation}
and the integration of the above equation leads to
\begin{equation}\label{53a}
U={\textsf{C}}\int^{r}_{0}\left[\frac{\sigma}{{\textsf{C}}}+\frac{4\pi}{E}\left({\zeta}-\frac{T^{\textsf{(GT)}}_{01}}{{\textsf{A}}{\textsf{B}}}\right)
\right]{\textsf{C}}'dr+h(t){\textsf{C}},
\end{equation}
where integration constant is represented by $h(t)$ and its value at
the boundary gives
\begin{equation}\label{54a}
U=\frac{U_{\Sigma}}{{\textsf{C}}_{\Sigma}}{\textsf{C}}-{\textsf{C}}\int^{r_{\Sigma}}_{r}
\left[\frac{\sigma}{{\textsf{C}}}+\frac{4\pi}{E}\left({\zeta}-\frac{T^{\textsf{(GT)}}_{01}}{{\textsf{A}}{\textsf{B}}}\right)
\right]{\textsf{C}}'dr.
\end{equation}
The factors responsible for the deviation of matter source from the
homologous mode are heat dissipation and shear scalar. To observe
the necessary condition for homologous evolution \cite{22a,25},
i.e., $U\sim {\textsf{C}}$ the integrand in the above equation must
cancel the effect of each other. Further, it yields
$U=h(t){\textsf{C}}$ where
$h(t)=\frac{U_{\Sigma}}{{\textsf{C}}_{\Sigma}}$. In $f(G,T)$
gravity, the homologous condition is of the form
\begin{equation}\label{56a}
\frac{\sigma}{{\textsf{C}}}+\frac{4\pi
{\textsf{B}}}{{\textsf{C}}'}\left({\zeta}-\frac{T^{\textsf{(GT)}}_{01}}{{\textsf{A}}{\textsf{B}}}\right)=0.
\end{equation}

\subsection{Homogeneous Expansion}

Another mode under discussion is homogeneous expansion for which the
required condition is $\Theta'=0$. In contrast to preceding mode,
the homogeneous contraction or expansion occurs when its rate is
independent of $r$. The combination of this condition on
Eq.\eqref{21a} becomes
\begin{equation}\label{57a}
4\pi\left({\zeta}-\frac{T^{\textsf{(GT)}}_{01}}{{\textsf{A}}{\textsf{B}}}\right)=-\frac{{\textsf{C}}'}{{\textsf{B}}}
\left[\frac{\sigma}{{\textsf{C}}}+\frac{1}{3}D_{{\textsf{C}}}(\sigma)\right].
\end{equation}
Utilizing the homologous condition in Eq.\eqref{57a} provides
$D_{{\textsf{C}}}(\sigma)=0$. The regularity conditions at the core
give rise to $\sigma=0$, and its implication in the above equation
yields
\begin{equation}\label{59a}
{\zeta}=\frac{T^{\textsf{(GT)}}_{01}}{{\textsf{A}}{\textsf{B}}}.
\end{equation}
This shows that the fluid incorporates the influence of heat
dissipation. In the framework of general relativity (GR), the matter
distribution was non-dissipative as well as shear-free during
homogeneous evolution.

\section{Kinematical and Dynamical Quantities}

Here, we investigate the behavior of some physical entities to
select the simplest evolutionary mode. For our convenience, we
assume that the metric coefficient ${\textsf{C}}$ is a separable
function of $t$ and $r$. Employing homologous condition \eqref{56a}
in Eq.\eqref{21a}, we obtain
\begin{equation}\label{62a}
\left(\Theta-\sigma\right)'=\left(\frac{3\dot
{\textsf{C}}}{{\textsf{A}}{\textsf{C}}}\right)'=0.
\end{equation}
This equation indicates the geodesic nature of matter configuration
as ${\textsf{A}}'=0 ~(a=0)$ implying that ${\textsf{A}}=1$ without
any loss of generality. Combining the values of $\sigma$ and
$\Theta$ together with the geodesic condition leads to
\begin{equation}\label{63a}
\Theta-\sigma=\frac{3\dot {\textsf{C}}}{{\textsf{C}}}.
\end{equation}
The homologous condition is recovered by taking the successive
derivatives of Eq.\eqref{63a}. One can observe that the necessary
and sufficient condition for the dynamical sphere to evolve
homologously is that the fluid must obey the geodesic path and
vice-versa. In GR, the unavailability of heat flux (${\zeta}=0$)
corresponds to a shear-free matter distribution, while in this
modified theory its effect can be seen as
\begin{equation}\label{64a}
\sigma=4\pi\frac{T^{\textsf{(GT)}}_{01}{\textsf{C}}}{{\textsf{C}}'}.
\end{equation}
If the matter source follows the homogeneous collapse with no
dissipation then we have $T^{\textsf{(GT)}}_{01}=0$, while the shear
scalar from Eq.\eqref{57a} becomes
\begin{equation}\label{65a}
\sigma=\frac{12 \pi}{{\textsf{C}}^3}\int
\frac{{\textsf{C}}^3T^{\textsf{(GT)}}_{01}}{{\textsf{A}}}dr+\frac{b
(t)}{{\textsf{C}}^3},
\end{equation}
where $b(t)$ stands for the integration function. Thus, one can say
that when we neglect the additional terms and heat dissipation, the
homogeneous expansion implies homologous condition as
$\sigma=0\Rightarrow U\sim {\textsf{C}}$. This result shows that
homologous evolution is the simplest mode. During collapse, the mass
and velocity undergoing homologous evolution are interlinked as
\begin{equation}\label{67a}
D_{T}U=\frac{\mathbf{s}^2}{2\textsf{C}^3}-\frac{m}{{\textsf{C}}^2}-4\pi\left({\mathfrak{p}}_{r}
-\frac{\mathbf{s}^2}{8\pi\textsf{C}^4}+\frac{T^{\textsf{(GT)}}_{11}}{{\textsf{B}}^2}\right){\textsf{C}}.
\end{equation}
In terms of ${\mathcal{Y}}_{TF}$, the above expression can be
obtained as
\begin{align}\nonumber
\frac{3D_{T}U}{{\textsf{C}}}&={\mathcal{Y}}_{TF}-\frac{\Pi}{2}f_{T}-\textsf{L}^{\textsf{(GT)}}+\frac{\mathbf{s}^2}{\textsf{C}^4}+4\pi\Pi^{\textsf{(GT)}}-4\pi
\bigg[\bigg({\mathfrak{u}}+\frac{\mathbf{s}^2}{8\pi\textsf{C}^4}+\frac{T^{\textsf{(GT)}}_{00}}{{\textsf{A}}^2}\bigg)
\\\label{68a}&+3\bigg({\mathfrak{p}}_{r}-\frac{\mathbf{s}^2}{8\pi\textsf{C}^4}
+\frac{T^{\textsf{(GT)}}_{11}}{{\textsf{B}}^2}\bigg)-2\Pi\bigg].
\end{align}
After some manipulations, we have
\begin{equation}\label{71a}
\frac{\ddot {\textsf{C}}}{{\textsf{C}}}-\frac{\ddot
{\textsf{B}}}{{\textsf{B}}}={\mathcal{Y}}_{TF}-\textsf{L}^{\textsf{(GT)}}+\frac{\mathbf{s}^2}{\textsf{C}^4}-\frac{\Pi}{2}f_{T}+4\pi\Pi^{\textsf{(GT)}}.
\end{equation}
This shows that charged spherical distribution gets complexity-free
when
\begin{equation}\label{73a}
\frac{\ddot
{\textsf{C}}}{{\textsf{C}}}-4\pi\Pi^{\textsf{(GT)}}-\frac{\mathbf{s}^2}{\textsf{C}^4}-\frac{\ddot
{\textsf{B}}}{{\textsf{B}}}+L^{\textsf{(GT)}}+\frac{\Pi}{2}f_{T}=0.
\end{equation}

Here, we find feasible solutions corresponding to non-dissipative
and dissipative scenarios by using zero complexity and homologous
conditions. The system under consideration is non-static and
modified field equations contain non-linear terms, thus, a linear
$f(G,T)$ model of the form $f(G,T)=\gamma G^n+\omega T$
\cite{23,23a} is chosen for the sake of simplicity. Moreover,
$\gamma$ and $\omega$ denote the real numbers and $n>0$, thus we can
take $\gamma=n=1$. There are three metric potentials
(${\textsf{A}}(t,r)$, ${\textsf{B}}(t,r)$, ${\textsf{C}}(t,r)$) to
be calculated in the current setup. As we have selected
${\textsf{A}}=1$, so we are left with only two unknowns. It means
that two equations are required to find the unknowns. For this
reason, the zero complexity and homologous conditions are utilized.
These constraints, in the case of non-dissipative fluid, are
calculated as
\begin{eqnarray}\nonumber
&&\frac{1}{{{\textsf{B}}^3 {\textsf{C}}^2(\omega +8 \pi
)}}\left[\left(-2 {\textsf{C}} \left(5 \ddot {\textsf{B}} \dot
{\textsf{C}}'+{\textsf{C}}'\dot {\textsf{B}} \right)+\dot
{\textsf{B}} {\textsf{C}}'\ddot {\textsf{C}}\right){\textsf{B}}+10
\dot {\textsf{B}} \ddot {\textsf{B}} {\textsf{C}}
{\textsf{C}}'\right.\\\label{77a}&&\left.+2 {\textsf{B}}^2 \dot
{\textsf{C}}' \left({\textsf{C}}-4 \ddot
{\textsf{C}}\right)\right]=0,
\end{eqnarray}
\begin{eqnarray}\nonumber
&&\frac{1}{2 {\textsf{B}}^4 {\textsf{C}}^4}\{2
{\textsf{B}}^3\dot{{\textsf{B}}} {\textsf{C}}^2 \dot{{\textsf{C}}}
\big(-4 \ddot{{\textsf{C}}}+{\textsf{C}}\big)+{\textsf{B}}
{\textsf{C}}^2 {\textsf{C}}^{'} \big(-2 {\textsf{B}}^{'}
{\textsf{C}}+\big(\dot{{\textsf{B}}}^2+3 \ddot{{\textsf{B}}}\big)
{\textsf{C}}^{'}\\\nonumber&&+32 \dot{{\textsf{B}}}
\dot{{\textsf{C}}'}^2\big)-4 \dot{{\textsf{B}}}^2 {\textsf{C}}^2
{{\textsf{C}}'}^2+{\textsf{B}}^4 \big({\mathbf{s}}^2-2
{\textsf{C}}^2 \big(\dot{{\textsf{C}}}^2+1\big)\big)+{\textsf{B}}^2
{\textsf{C}}^2 \big(2 {{\textsf{C}}'}^2+16
\dot{{\textsf{C}}'}^2\\\label{78a}&&-2 {\textsf{C}}
\dot{{\textsf{C}}}+{\mathbf{s}}^2\big)\}=0.
\end{eqnarray}
For the second case (dissipative fluid), the complexity-free
condition remains the same as for non-dissipative source while the
homologous condition turns out to be
\begin{eqnarray}\nonumber
{\zeta}&=&\frac{1}{16\pi^2 {\textsf{C}}^4{\textsf{B}}^5 (\omega +1)
}\big[{\textsf{C}}'\big(-{\textsf{B}} \dot {\textsf{C}}+\dot
{\textsf{B}}{\textsf{C}}\big) \big({\textsf{B}} \big({\textsf{C}}
\big(40 \pi\dot {\textsf{C}}' \ddot {\textsf{B}}+\dot {\textsf{B}}
{\textsf{C}}' \big)-4 \pi\ddot {\textsf{C}} \dot {\textsf{B}}
{\textsf{C}}'\big)\big.\big.\big.\\\label{79a}\big.\big.\big.&-&40
\pi \dot {\textsf{B}} \ddot {\textsf{B}} {\textsf{C}}
{\textsf{C}}'+{\textsf{B}}^2 \big(-\dot {\textsf{C}}'\big)
\big({\textsf{C}}-32 \pi \ddot {\textsf{C}}\big)\big)\big].
\end{eqnarray}

\section{Stability of Zero Complexity Condition}

It is possible that the system has zero complexity at first and then
becomes complicated due to the presence of some factors as it
evolves. These factors are determined by working on the evolution
equation for the scalar $\mathcal{Y}_{TF}$ following the technique
\cite{24a} together Eqs.\eqref{13a}, \eqref{43a} and \eqref{45a} as
\begin{eqnarray}\nonumber
&&-4\pi\left({\mathfrak{p}}_{r}+\frac{T^{11\textsf{(GT)}}}{{\textsf{B}}^2}+{\mathfrak{u}}+T^{00\textsf{(GT)}}\right)\sigma
+\frac{\mathbf{s}^2}{\textsf{C}^4}+\frac{\mathbf{s}^2\mathbf{s}'}{\textsf{C}^4\textsf{B}}
-\frac{4\pi}{{\textsf{B}}}\left[{\zeta}'-(\frac{T^{00\textsf{(GT)}}}{{\textsf{B}}^2})'{\textsf{B}}
\right.\\\nonumber&&\left.-({\zeta}-\frac{T^{01\textsf{(GT)}}}{{\textsf{B}}^2})\frac{{\textsf{C}}'}{{\textsf{C}}}\right]
-\dot
{\mathcal{Y}}_{TF}+\frac{3\mathbf{s}\dot{\mathbf{s}}}{\textsf{C}^4}-\dot
{\textsf{L}}^{\textsf{(GT)}}-8\pi\dot
\Pi-4\pi\dot\Pi^{\textsf{(GT)}}-4\pi
\left(T^{\textsf{(GT)}}_{00}\right)^.\\\nonumber&&-3\frac{\dot
{\textsf{C}}}{{\textsf{C}}}\left({\mathcal{Y}}_{TF}-\textsf{L}^{\textsf{(GT)}}\right)-12\pi\big(
T^{00\textsf{(GT)}}+\frac{\mathbf{s}^2}{8\pi\textsf{C}^4}\big)\frac{\dot
{\textsf{C}}}{{\textsf{C}}}-4\pi \texttt{Z}_{1}+\frac{4\pi
T^{00\textsf{(GT)}}{\textsf{B}}'}{{\textsf{B}}^2}\\\label{87a}&&+\left(12\pi
T^{\textsf{(GT)}}_{00}-12\pi\Pi^{\textsf{(GT)}}+\frac{12\pi
T^{\textsf{(GT)}}_{11}}{{\textsf{B}}^2}\right)\frac{\dot
{\textsf{C}}}{{\textsf{C}}}-\frac{16\pi\Pi^{\textsf{(tot)}}\dot
{\textsf{C}}}{{\textsf{C}}}-\frac{5\mathbf{s}^2\dot{\textsf{C}}}{2\textsf{C}^5}
=0.
\end{eqnarray}
For the non-dissipative source, we substitute
${\mathcal{Y}}_{TF}=\Pi^{\textsf{(tot)}}={\zeta}=\sigma=0$ in the
above equation at the initial time ($t=0$), so that it becomes
\begin{eqnarray*}\nonumber
&&-\dot {\mathcal{Y}}_{TF}-4\pi
\left(T^{\textsf{(GT)}}_{00}\right)^.-\dot
{\textsf{L}}^{\textsf{(GT)}}+4\pi(\frac{T^{00\textsf{(GT)}}}{{\textsf{B}}^2})'-4\pi\dot\Pi^{\textsf{(GT)}}-8\pi\dot
\Pi+3\textsf{L}^{\textsf{(GT)}}\\\nonumber&&\times\frac{\dot
{\textsf{C}}}{{\textsf{C}}}-4\pi \texttt{Z}_{1}-12\pi\big(
T^{00\textsf{(GT)}}+\frac{\mathbf{s}^2}{8\pi\textsf{C}^4}\big)\frac{\dot
{\textsf{C}}}{{\textsf{C}}}+\frac{4\pi
T^{00\textsf{(GT)}}{\textsf{B}}'}{{\textsf{B}}^2}+\big(12\pi
T^{\textsf{(GT)}}_{00}+\frac{12\pi
T^{\textsf{(GT)}}_{11}}{{\textsf{B}}^2}\big)\\&&\times\frac{\dot
{\textsf{C}}}{{\textsf{C}}}+\frac{\mathbf{s}^2}{\textsf{C}^4}+\frac{\mathbf{s}^2\mathbf{s}'}{\textsf{C}^4\textsf{B}}
+\frac{3\mathbf{s}\dot{\mathbf{s}}}{\textsf{C}^4}-\frac{5\mathbf{s}^2\dot{\textsf{C}}}{2\textsf{C}^5}=0.
\end{eqnarray*}
Making use of the above constraint with the derivative of
Eq.\eqref{47a} at $t=0$ provides
\begin{eqnarray}\nonumber
&&\frac{4\pi}{{\textsf{C}}^3}\int
^{r}_{0}[({\mathfrak{u}}+\frac{\mathbf{s}^2}{8\pi\textsf{C}^4}+\frac{T^{\textsf{(GT)}}_{00}}{{\textsf{A}}^2})^.]'=
4\pi(\frac{T^{00\textsf{(GT)}}}{{\textsf{B}}^2})'-4\pi
\left(T^{\textsf{(GT)}}_{00}\right)^.+\frac{4\pi
T^{00\textsf{(GT)}}{\textsf{B}}'}{{\textsf{B}}^2}\\\nonumber&&-12\pi\big(
T^{00\textsf{(GT)}}+\frac{\mathbf{s}^2}{8\pi\textsf{C}^4}\big)\frac{\dot
{\textsf{C}}}{{\textsf{C}}}+3L^{\textsf{(GT)}}\frac{\dot
{\textsf{C}}}{{\textsf{C}}}-4\pi \texttt{Z}_{1}+(12\pi
T^{\textsf{(GT)}}_{00}+\frac{12\pi
T^{\textsf{(GT)}}_{11}}{{\textsf{B}}^2})\\\label{90b}&&\times\frac{\dot
{\textsf{C}}}{{\textsf{C}}}+\frac{\mathbf{s}^2}{\textsf{C}^4}+\frac{\mathbf{s}^2\mathbf{s}'}{\textsf{C}^4\textsf{B}}
+\frac{4\mathbf{s}\dot{\mathbf{s}}}{\textsf{C}^4}-\frac{\mathbf{s}^2\dot{\textsf{C}}}{2\textsf{C}^5}.
\end{eqnarray}

The stability of the complexity-free condition is associated with
the pressure and energy density. The above equation demonstrates
that the contributions of pressure anisotropy, charge, modified
terms and energy density inhomogeneity in the inner matter source
forced the system to deviate from the stable position. This depicts
that the stability of the system is disturbed by the presence of
electromagnetic field. Further, using the conditions
${\mathcal{Y}}_{TF}=\sigma=0$ in conjunction with the derivative of
Eq.\eqref{87a} with respect to $t$ yields
\begin{align}\nonumber
&-\ddot
{\mathcal{Y}}_{TF}-4\pi\ddot\Pi^{\textsf{(GT)}}+\left[4\pi\left(\frac{T^{00\textsf{(GT)}}}{{\textsf{B}}^2}\right)'\right]^.
-\ddot
{\textsf{L}}^{\textsf{(GT)}}-4\pi\left(T^{\textsf{(GT)}}_{00}\right)^{..}+\dot
{\textsf{L}}^{\textsf{(GT)}}-4\pi \dot
Z_{1}\\\nonumber&-12\pi\big\{\big(
\frac{\mathbf{s}^2}{8\pi\textsf{C}^4}+T^{00\textsf{(GT)}}\big)\frac{\dot
{\textsf{C}}}{{\textsf{C}}}\big\}^.+\frac{3\dot
{\textsf{C}}}{{\textsf{C}}}[4\pi\left(T^{\textsf{(GT)}}_{00}\right)^{.}-4\pi\left(\frac{T^{00\textsf{(GT)}}}{{\textsf{B}}^2}\right)'
+4\pi\dot\Pi^{\textsf{(GT)}}\\\nonumber&+4\pi
\texttt{Z}_{1}+\frac{3\mathbf{s}\dot{\mathbf{s}}}{\textsf{C}^4}-4\pi\left(\frac{{\textsf{B}}'T^{00\textsf{(GT)}}}{{\textsf{B}}^2}\right)-3L^{\textsf{(GT)}}\frac{\dot
{\textsf{C}}}{{\textsf{C}}}+12\pi\left(\frac{T^{00\textsf{(GT)}}\dot
{\textsf{C}}}{{\textsf{C}}}\right)+8\pi\dot\Pi
-\frac{5\mathbf{s}^2\dot{\textsf{C}}}{2\textsf{C}^5}\\\nonumber&-\left(12\pi
T^{\textsf{(GT)}}_{00}+\frac{12\pi
T^{\textsf{(GT)}}_{11}}{{\textsf{B}}^2}\right)\frac{\dot
{\textsf{C}}}{{\textsf{C}}}+\frac{\mathbf{s}^2}{\textsf{C}^4}+\frac{\mathbf{s}^2\mathbf{s}'}{\textsf{C}^4\textsf{B}}
]+3\left(\frac{L^{\textsf{(GT)}}\dot
{\textsf{C}}}{{\textsf{C}}}\right)^.-\frac{16\pi\Pi^{\textsf{(tot)}}\dot
{\textsf{C}}}{{\textsf{C}}}-8\pi\ddot\Pi\\\nonumber&+\left[(12\pi
T^{\textsf{(GT)}}_{00}+\frac{12\pi
T^{\textsf{(GT)}}_{11}}{{\textsf{B}}^2})\frac{\dot
{\textsf{C}}}{{\textsf{C}}}\right]^.+4\pi\left(\frac{{\textsf{B}}'T^{00\textsf{(GT)}}}{{\textsf{B}}^2}\right)^.
+\big(\frac{\mathbf{s}^2}{\textsf{C}^4}+\frac{\mathbf{s}^2\mathbf{s}'}{\textsf{C}^4\textsf{B}}+\frac{3\mathbf{s}\dot{\mathbf{s}}}{\textsf{C}^4}
-\frac{5\mathbf{s}^2\dot{\textsf{C}}}{2\textsf{C}^5}\big)^.=0.\\\label{89a}
\end{align}
It is worth mentioning here that for the dissipative case, the heat
flux works as an additional factor in disturbing the vanishing
complexity constraint.

\section{Concluding Remarks}

In this paper, we have checked the influence of non-minimal theory
to calculate the complexity of charged dynamical structure. The
interior region was filled with energy density inhomogeneity,
anisotropic pressure, heat flux and charge. Using Herrera's method,
we have divided the Riemann tensor into four scalars, each of which
is associated with a certain set of physical attributes. Among all
the four scalars, we have considered ${\mathcal{Y}}_{TF}$ to produce
complexity for the reasons stated below.
\begin{enumerate}
\item In GR and $f(G,T)$ gravity, this factor has already been worked
as complexity factor for the static spacetime. Thus, we can reobtain
this complexity-producing scalar for the static spherical system
from the non-static geometry by using Eq.\eqref{43a}.
\item All the physical parameters (anisotropic pressure, energy density
inhomogeneity, heat flux, charge and correction terms) are
incorporated in this factor that might increase the complexity of
the system.
\end{enumerate}
We have looked at two different types of evolutionary patterns,
homologous and homogenous. By using the criterion of zero complexity
and selecting homologous mode as the most basic pattern of
evolution, we are able to obtain the results for both dissipative
and non-dissipative scenarios. The variables that lead the system to
depart from zero complexity criterion throughout the evolution
process have also been covered.

The fundamental properties of the charged complex sphere are
strongly influenced by the state determinants and additional
curvature factors in $f(G,T)$ gravity. As a result, the charged
dynamical sphere becomes complicated due to the inclusion of higher
order curvature factors. Since the homologous fluid is assumed to be
geodesic in nature ($\textsf{A}=1$), this mode was proposed as the
most basic pattern of evolution. Further, the homologous condition
contains the extra terms of this theory. In GR, the conditions
${\mathfrak{u}}=\Pi={\zeta}=0$ are responsible for the formation of
a complex free system (${\mathcal{Y}}_{TF}=0$), whereas in this
theory, the structure corresponds to complexity-free only when the
constraint
$L^{\textsf{(GT)}}-\frac{\mathbf{s}^2}{\textsf{C}^4}+\frac{\Pi}{2}f_{T}-4\pi\Pi^{\textsf{(GT)}}=0$
and the aforementioned factors are satisfied. We have seen that the
presence of correction terms causes the shear to be established even
in the non-dissipative case. We have employed the modified
complexity free and homologous conditions to assess the feasible
solutions representing dissipative and non-dissipative scenarios.
The stability criterion for the vanishing complexity condition has
also been analyzed. Finally, we have addressed the factors
responsible for the deviation of the system from zero complexity. We
have also found that the contribution of charge and extra curvature
terms of this modified theory make the system more complex. It is
worth mentioning here that the constraint $f(G,T)=0$ reduces all the
obtained results to GR.

\section*{Appendix A}
\renewcommand{\theequation}{A\arabic{equation}}
\setcounter{equation}{0} The additional terms present in the field
equations of $f(G,T)$ theory are described as
\begin{eqnarray}\nonumber
T^{\textsf{(GT)}}_{00}&=&\frac{1}{8\pi}\left[\left({\mathfrak{u}}+{\mathfrak{p}}\right){\textsf{A}}^2f_{T}-\frac{{\textsf{A}}^2}{2}f
+\left(\frac{8\dot{{\textsf{B}}}\dot{{\textsf{C}}}\ddot{{\textsf{C}}}}{{\textsf{A}}^2{\textsf{B}}
{\textsf{C}}^2}-\frac{16{\textsf{A}}'\dot {\textsf{C}}\dot
{\textsf{C}}'}{{\textsf{A}}{\textsf{B}}^2{\textsf{C}}^2}+\frac{8{\textsf{A}}{\textsf{C}}'{\textsf{A}}'
{\textsf{C}}''}{{\textsf{B}}^4{\textsf{C}}^2}\right.\right.\\\nonumber
&+&\left.\left. \frac{4\ddot
{\textsf{B}}}{{\textsf{B}}{\textsf{C}}^2}+\frac{8{\textsf{A}}'\dot
{\textsf{B}}\dot
{\textsf{C}}{\textsf{C}}'}{{\textsf{A}}{\textsf{B}}^3{\textsf{C}}^2}-\frac{8\dot
{\textsf{A}}{\textsf{B}}'{\textsf{C}}'\dot
{\textsf{C}}}{{\textsf{A}}{\textsf{B}}^3{\textsf{C}}^2}+\frac{8\dot
{\textsf{A}}\dot
{\textsf{C}}{\textsf{C}}''}{{\textsf{A}}{\textsf{B}}^2{\textsf{C}}^2}+\frac{4\dot
{\textsf{C}}^2{\textsf{A}}'}{{\textsf{A}}{\textsf{C}}^2}-\frac{12\dot
{\textsf{A}}\dot {\textsf{B}}\dot
{\textsf{C}}^2}{{\textsf{A}}^3{\textsf{B}}{\textsf{C}}^2}\right.\right.\\\nonumber
&\times&\left.\left.\frac{{\textsf{B}}'}{{\textsf{B}}^3}+\frac{4{\dot
{\textsf{A}}\textsf{C}}'^2\dot
{\textsf{B}}}{{\textsf{A}}{\textsf{B}}^3{\textsf{C}}^2}-\frac{12{\textsf{A}}{\textsf{A}}'{\textsf{B}}'{\textsf{C}}'^2}
{{\textsf{B}}^5{\textsf{C}}^2}+\frac{4{\textsf{A}}{\textsf{C}}'^2{\textsf{A}}''}{{\textsf{B}}^4{\textsf{C}}^2}+\frac{4\dot
{\textsf{C}}^2\ddot {\textsf{B}}
}{{\textsf{A}}^2{\textsf{B}}{\textsf{C}}^2}+\frac{8\dot
{\textsf{C}}'^2}{{\textsf{B}}^2{\textsf{C}}^2}\right.\right.\\\nonumber
&+&\left.\left.\frac{8{\textsf{B}}'\ddot
{\textsf{C}}{\textsf{C}}'}{{\textsf{B}}^3{\textsf{C}}^2}-\frac{16\dot
{\textsf{B}}{\textsf{C}}'\dot
{\textsf{C}}'}{{\textsf{C}}^2{\textsf{B}}^3}-\frac{4\dot
{\textsf{A}}\dot
{\textsf{B}}}{{\textsf{A}}{\textsf{C}}^2{\textsf{B}}}-\frac{4\dot
{\textsf{C}}^2{\textsf{A}}''}{{\textsf{B}}^2{\textsf{A}}{\textsf{C}}^2}+\frac{8\dot
{\textsf{C}}^2{\textsf{A}}'^2}{{\textsf{B}}^2{\textsf{A}}^2{\textsf{C}}^2}+\frac{8{\textsf{C}}'^2\dot
{\textsf{B}}^2}{{\textsf{C}}^2{\textsf{B}}^4}\right.\right.\\\nonumber
&-&\left.\left.\frac{8\ddot
{\textsf{C}}{\textsf{C}}''}{{\textsf{C}}^2{\textsf{B}}^2}-\frac{4{\textsf{A}}{\textsf{A}}''}{{\textsf{C}}^2{\textsf{B}}^2}
+\frac{{\textsf{B}}'{\textsf{A}}{\textsf{A}}'}{{\textsf{B}}^3{\textsf{C}}^2}
-\frac{4{\textsf{C}}'^2\ddot
{\textsf{B}}}{{\textsf{B}}^3{\textsf{C}}^2}\right)f_{G}+\left(\frac{8\dot
{\textsf{C}}{\textsf{C}}''}{{\textsf{B}}^2{\textsf{C}}^2}-\frac{8{\textsf{B}}'{\textsf{C}}'\dot
{\textsf{C}}}{{\textsf{B}}^3{\textsf{C}}^2}\right.\right.\\\nonumber
&-&\left.\left.\frac{4\dot
{\textsf{B}}}{{\textsf{B}}{\textsf{C}}^2}-\frac{12\dot
{\textsf{B}}\dot
{\textsf{C}}^2}{{\textsf{A}}^2{\textsf{B}}{\textsf{C}}^2}+\frac{4\dot
{\textsf{B}}{\textsf{C}}'^2}{{\textsf{B}}^3{\textsf{C}}^2}\right)\dot
f_{G}+\left(\frac{8\dot {\textsf{B}} \dot
{\textsf{C}}{\textsf{C}}'}{{\textsf{B}}^3{\textsf{C}}^2}-\frac{4{\textsf{B}}'\dot
{\textsf{C}}^2}{{\textsf{B}}^3{\textsf{C}}^2}-\frac{4{\textsf{A}}^2{\textsf{B}}'}{{\textsf{B}}^3{\textsf{C}}^2}\right.\right.
\\\nonumber&+&\left.\left.\frac{12{\textsf{A}}^2{\textsf{B}}'{\textsf{C}}'^2}{{\textsf{B}}^5{\textsf{C}}^2}
-\frac{8{\textsf{A}}^2{\textsf{C}}'{\textsf{C}}''}{{\textsf{B}}^4{\textsf{C}}^2}\right)f'_{G}
+\left(\frac{4\dot
{\textsf{C}}^2}{{\textsf{B}}^2{\textsf{C}}^2}+\frac{4{\textsf{A}}^2}{{\textsf{B}}^2{\textsf{C}}^2}
-\frac{4{\textsf{A}}^2{\textsf{C}}'^2}{{\textsf{B}}^4{\textsf{C}}^2}\right)\right.\\\label{100}
&\times&\left.f''_{G}\right],
\\\nonumber
T^{\textsf{(GT)}}_{01}&=&\frac{1}{8\pi}\left[-{\zeta}{\textsf{A}}{\textsf{B}}f_{T}+\left(\frac{10{\textsf{A}}'{\textsf{B}}'{\textsf{C}}'\dot
{\textsf{B}}}{{\textsf{A}}{\textsf{B}}^4{\textsf{C}}}-\frac{10\dot
{\textsf{A}}\dot {\textsf{B}}\dot
{\textsf{C}}{\textsf{A}}'}{{\textsf{A}}^4{\textsf{B}}{\textsf{C}}}-\frac{8\dot
{\textsf{A}}\dot {\textsf{B}}\dot
{\textsf{C}}'}{{\textsf{A}}^3{\textsf{B}}{\textsf{C}}^2}-\frac{8{\textsf{A}}'
}{{\textsf{A}}{\textsf{B}}^3}\right.\right.\\\nonumber
&\times&\left.\left.\frac{\dot
{\textsf{B}}{\textsf{C}}'^2}{{\textsf{C}}^2}-\frac{10\dot
{\textsf{A}}\dot
{\textsf{B}}^2{\textsf{C}}'}{{\textsf{A}}^3{\textsf{B}}^2{\textsf{C}}}-\frac{8{\textsf{A}}'^2{\textsf{C}}'\dot
{\textsf{C}}}{{\textsf{A}}^2{\textsf{B}}^2{\textsf{C}}^2}+\frac{10{\textsf{A}}'^2{\textsf{B}}'\dot
{\textsf{C}}}{{\textsf{A}}^2{\textsf{B}}^3{\textsf{C}}}+\frac{\dot
{\textsf{B}}\ddot
{\textsf{C}}{\textsf{C}}'}{{\textsf{A}}^2{\textsf{B}}{\textsf{C}}^2}+\frac{10\dot
{\textsf{B}} \ddot
{\textsf{B}}}{{\textsf{A}}^2{\textsf{B}}^2}\right.\right.\\\nonumber
&\times&\left.\left.\frac{{\textsf{C}}'}{{\textsf{C}}}+\frac{10{\textsf{A}}'\ddot
{\textsf{B}}\dot
{\textsf{C}}}{{\textsf{A}}^3{\textsf{B}}{\textsf{C}}}-\frac{10{\textsf{A}}''{\textsf{A}}'\dot
{\textsf{C}}}{{\textsf{A}}^2{\textsf{B}}^2{\textsf{C}}}-\frac{10{\textsf{A}}''{\textsf{C}}'\dot
{\textsf{B}}}{{\textsf{A}}{\textsf{B}}^3{\textsf{C}}}+\frac{8{\textsf{A}}'{\textsf{C}}'}{{\textsf{A}}{\textsf{B}}^2}+\frac{\dot
{\textsf{A}}\dot {\textsf{B}}\dot
{\textsf{C}}'}{{\textsf{A}}^3{\textsf{B}}{\textsf{C}}}-10\right.\right.\\\nonumber
&-&\left.\left.\frac{{\textsf{A}}'\dot
{\textsf{C}}'{\textsf{B}}'\dot
{\textsf{C}}'}{{\textsf{A}}{\textsf{B}}^3{\textsf{C}}{\textsf{C}}^2}+\frac{8{\textsf{A}}'\dot
{\textsf{C}}\ddot
{\textsf{C}}}{{\textsf{A}}^3{\textsf{C}}^2}-\frac{8{\textsf{A}}'\dot
{\textsf{A}}\dot
{\textsf{C}}^2}{{\textsf{A}}^4{\textsf{C}}^2}+\frac{8\dot
{\textsf{A}}\dot {\textsf{C}}\dot
{\textsf{C}}'}{{\textsf{A}}^3{\textsf{C}}^2}-\frac{10\ddot
{\textsf{B}}\dot
{\textsf{C}}'}{{\textsf{A}}^2{\textsf{B}}{\textsf{C}}}+\frac{10{\textsf{A}}''\dot
{\textsf{C}}'}{{\textsf{A}}{\textsf{B}}^2{\textsf{C}}}\right.\right.\\\nonumber&-&\left.\left.\frac{8\ddot
{\textsf{C}}\dot
{\textsf{C}}'}{{\textsf{A}}^2{\textsf{C}}^2}\right)f_{G}+\left(-\frac{8\dot
{\textsf{A}}\dot
{\textsf{B}}{\textsf{A}}'}{{\textsf{A}}^4{\textsf{B}}}-\frac{4{\textsf{A}}'}{{\textsf{A}}{\textsf{C}}^2}+\frac{8\dot
{\textsf{C}}\dot
{\textsf{C}}'}{{\textsf{A}}^2{\textsf{C}}^2}+\frac{4{\textsf{A}}'{\textsf{C}}'^2}{{\textsf{A}}{\textsf{B}}^2{\textsf{C}}^2}
+\frac{8{\textsf{A}}'^2{\textsf{B}}'}{{\textsf{A}}^2{\textsf{B}}^3}\right.\right.\\\nonumber&-&\left.\left.\frac{12{\textsf{A}}'\dot
{\textsf{C}}^2}{{\textsf{A}}^3{\textsf{C}}^2}-\frac{8{\textsf{A}}'{\textsf{A}}''}{{\textsf{A}}^2{\textsf{B}}^2}+\frac{8{\textsf{A}}'\ddot
{\textsf{B}}}{{\textsf{A}}^3{\textsf{B}}}\right)\dot
f_{G}\left(\frac{8{\textsf{A}}'{\textsf{C}}'\dot
{\textsf{C}}}{{\textsf{A}}{\textsf{B}}^2{\textsf{C}}^2}+\frac{8{\textsf{A}}'{\textsf{B}}'\dot
{\textsf{B}}}{{\textsf{A}}{\textsf{B}}^4}-\frac{4\dot
{\textsf{B}}\dot
{\textsf{C}}^2}{{\textsf{A}}^2{\textsf{B}}{\textsf{C}}^2}\right.\right.\\\nonumber&-&\left.\left.\frac{8\dot
{\textsf{A}}\dot
{\textsf{B}}^2}{{\textsf{A}}^3{\textsf{B}}^2}+\frac{12\dot
{\textsf{B}}{\textsf{C}}'^2}{{\textsf{B}}^3{\textsf{C}}^2}+\frac{8\ddot
{\textsf{B}}\dot
{\textsf{B}}}{{\textsf{A}}^2{\textsf{B}}^2}-\frac{8{\textsf{A}}''\dot
{\textsf{B}}}{{\textsf{A}}{\textsf{B}}^3}-\frac{8{\textsf{C}}'\dot
{\textsf{C}}'}{{\textsf{B}}^2{\textsf{C}}^2}-\frac{4\dot
{\textsf{B}}}{{\textsf{B}}{\textsf{C}}^2}\right)f'_{G}+\left(4\right.\right.
\\\label{100a}&\times&\left.\left.\frac{\dot
{\textsf{C}}^2}{{\textsf{A}}^2{\textsf{C}}^2}+\frac{4}{{\textsf{C}}^2}-\frac{4{\textsf{C}}'^2}{{\textsf{B}}^2{\textsf{C}}^2}
+\frac{8{\textsf{A}}''}{{\textsf{A}}{\textsf{B}}^2}-\frac{8\ddot
{\textsf{B}}}{{\textsf{A}}^2{\textsf{B}}}-\frac{8{\textsf{A}}'{\textsf{B}}'}{{\textsf{A}}{\textsf{B}}^3}+\frac{8\dot
{\textsf{A}}\dot
{\textsf{B}}}{{\textsf{A}}^3{\textsf{B}}}\right)\dot
f'_{G}\right],\\\nonumber
T^{\textsf{(GT)}}_{11}&=&\frac{1}{8\pi}\left[\frac{2}{3}\Pi
{\textsf{B}}^2 f_{T}+\frac{{\textsf{B}}^2}{2}f+\left(\frac{32\dot
{\textsf{B}}{\textsf{C}}' \dot
{\textsf{C}}'^2}{{\textsf{A}}^2{\textsf{B}}{\textsf{C}}^2}-\frac{8{\textsf{A}}'{\textsf{C}}'{\textsf{C}}''}{{\textsf{A}}{\textsf{B}}^2{\textsf{C}}^2}
-\frac{8{\textsf{B}}\dot {\textsf{B}}\dot {\textsf{C}}\ddot
{\textsf{C}}}{{\textsf{A}}^2{\textsf{C}}^4}-8\right.\right.\\\nonumber
&\times&\left.\left.\frac{{\textsf{B}}'\ddot
{\textsf{C}}{\textsf{C}}'}{{\textsf{B}}{\textsf{C}}^2{\textsf{A}}^2}+\frac{\ddot
{\textsf{B}}{\textsf{C}}'^2}{{\textsf{A}}^2{\textsf{B}}{\textsf{C}}^2}-\frac{4{\textsf{A}}''{\textsf{C}}'^2}{{\textsf{A}}{\textsf{B}}^2{\textsf{C}}^2}
-\frac{16\dot
{\textsf{C}}'^2}{{\textsf{A}}^2{\textsf{C}}^2}-\frac{8\dot
{\textsf{A}}\dot
{\textsf{C}}{\textsf{C}}''}{{\textsf{A}}^3{\textsf{C}}^2}+\frac{32{\textsf{A}}'\dot
{\textsf{C}}\dot
{\textsf{C}}'}{{\textsf{A}}^3{\textsf{C}}^2}\right.\right.\\\nonumber
&-&\left.\left.\frac{4{\textsf{B}}\ddot {\textsf{B}}\dot
{\textsf{C}}^2}{{\textsf{A}}^4{\textsf{C}}^2}-\frac{\dot
{\textsf{B}}^2{\textsf{C}}'^2}{{\textsf{A}}^2{\textsf{B}}^2{\textsf{C}}^2}+\frac{8\ddot
{\textsf{C}}{\textsf{C}}''}{{\textsf{A}}^2{\textsf{C}}^2}+\frac{4{\textsf{B}}\dot
{\textsf{B}}\dot
{\textsf{A}}}{{\textsf{A}}^3{\textsf{C}}^2}-\frac{4{\textsf{A}}'{\textsf{B}}'}{{\textsf{A}}{\textsf{B}}{\textsf{C}}^2}+\frac{4{\textsf{A}}''\dot
{\textsf{C}}^2}{{\textsf{A}}^3{\textsf{C}}^2}-24\right.\right.\\\nonumber
&\times&\left.\left.\frac{{\textsf{A}}'\dot {\textsf{B}}\dot
{\textsf{C}}{\textsf{C}}'}{{\textsf{A}}^3{\textsf{B}}{\textsf{C}}^2}-\frac{4{\textsf{B}}\ddot
{\textsf{B}}}{{\textsf{A}}^2{\textsf{C}}^2}+\frac{8\dot
{\textsf{A}}\dot
{\textsf{C}}{\textsf{B}}'{\textsf{C}}'}{{\textsf{A}}^3{\textsf{B}}{\textsf{C}}^2}-\frac{16{\textsf{A}}'^2\dot
{\textsf{C}}^2}{{\textsf{A}}^4{\textsf{C}}^2}+\frac{4{\textsf{A}}''}{{\textsf{A}}{\textsf{C}}^2}
+\frac{12{\textsf{A}}'{\textsf{C}}'^2{\textsf{B}}'}{{\textsf{A}}{\textsf{B}}^3
{\textsf{C}}^2}\right.\right.\\\nonumber
&-&\left.\left.\frac{4{\textsf{A}}'{\textsf{B}}'\dot
{\textsf{C}}^2}{{\textsf{A}}^3{\textsf{B}}{\textsf{C}}^2}-\frac{4\dot
{\textsf{A}}\dot
{\textsf{B}}{\textsf{C}}'^2}{{\textsf{A}}^3{\textsf{B}}{\textsf{C}}^2}\right)f_{G}+\left(\frac{8{\textsf{B}}^2\dot
{\textsf{C}}\ddot
{\textsf{C}}}{{\textsf{A}}^4{\textsf{C}}^2}-\frac{12{\textsf{B}}^2\dot
{\textsf{A}}\dot
{\textsf{C}}^2}{{\textsf{A}}^5{\textsf{C}}^2}+\frac{4\dot
{\textsf{A}}{\textsf{C}}'^2}{{\textsf{A}}^3{\textsf{C}}^2}\right.\right.\\\nonumber
&-&\left.\left.\frac{8{\textsf{A}}'\dot
{\textsf{C}}{\textsf{C}}'}{{\textsf{A}}^3{\textsf{C}}^2}-\frac{4{\textsf{B}}^2\dot
{\textsf{A}}}{{\textsf{A}}^3{\textsf{C}}^2}\right)\dot
f_{G}+\left(\frac{4{\textsf{B}}^2\dot
{\textsf{C}}^2}{{\textsf{A}}^4{\textsf{C}}^2}\frac{4{\textsf{B}}^2}{{\textsf{A}}^2{\textsf{C}}^2}
-\frac{4{\textsf{C}}'^2}{{\textsf{A}}^2{\textsf{C}}^2}\right)\ddot
f_{G}+\left(\frac{8\dot
{\textsf{C}}}{{\textsf{A}}^3}\right.\right.\\\label{100b}
&\times&\left.\left.\frac{\dot
{\textsf{A}}{\textsf{C}}'}{{\textsf{C}}^2}-\frac{4\dot
{\textsf{A}}'{\textsf{C}}^2}{{\textsf{A}}^3{\textsf{C}}^2}-\frac{4{\textsf{A}}'}{{\textsf{A}}{\textsf{C}}^2}
+\frac{12{\textsf{C}}'^2{\textsf{A}}'}{{\textsf{A}}{\textsf{B}}^2{\textsf{C}}^2}-\frac{\ddot
{\textsf{C}}{\textsf{C}}'}{{\textsf{A}}^2{\textsf{C}}^2}\right)f'_{G}\right],\\\nonumber
T^{\textsf{(GT)}}_{22}&=&\frac{1}{8\pi}\left[\frac{ -{\textsf{C}}^2
}{3}\Pi f_{T}+\frac{{\textsf{C}}^2}{2}f+\left(\frac{-4\dot
{\textsf{C}}^2}{{\textsf{A}}^2{\textsf{B}}^2}+\frac{4{\textsf{A}}''}{{\textsf{A}}{\textsf{B}}^2}+\frac{8\dot
{\textsf{A}}\dot
{\textsf{C}}{\textsf{B}}'{\textsf{C}}'}{{\textsf{A}}^3{\textsf{B}}^3}-\frac{4\ddot
{\textsf{B}}}{{\textsf{A}}^2{\textsf{B}}}-4\right.\right.\\\nonumber&\times&\left.\left.\frac{\dot
{\textsf{A}}\dot
{\textsf{B}}{\textsf{C}}'^2}{{\textsf{B}}^3{\textsf{A}}^3}-\frac{4{\textsf{A}}'{\textsf{B}}'\dot
{\textsf{C}}^2}{{\textsf{B}}^3{\textsf{A}}^3}+\frac{12\dot
{\textsf{B}}\dot {\textsf{A}}\dot
{\textsf{C}}^2}{{\textsf{B}}{\textsf{A}}^5}+\frac{12{{\textsf{B}}'\textsf{A}}'{\textsf{C}}'^2}{{\textsf{A}}{\textsf{B}}^5}
-\frac{8{\textsf{A}}'\textsf{C}
'{\textsf{C}}''}{{\textsf{B}}^4{\textsf{A}}}-\frac{8\dot
{\textsf{A}}\dot
{\textsf{C}}}{{\textsf{A}}^3}\right.\right.\\\nonumber&\times&\left.\left.\frac{{\textsf{C}}''}{{\textsf{B}}^2}
-\frac{4{\textsf{A}}'{\textsf{B}}'}{{\textsf{A}}{\textsf{B}}^3}-\frac{8\dot
{\textsf{B}}\dot {\textsf{C}} \ddot
{\textsf{C}}}{{\textsf{A}}^4{\textsf{B}}}-\frac{8{\textsf{C}}'{\textsf{B}}'\ddot
{\textsf{C}}}{{\textsf{B}}^3{\textsf{A}}^2}+\frac{8{\textsf{A}}'\dot
{\textsf{C}}\dot
{\textsf{C}}'}{{\textsf{A}}^3{\textsf{B}}^2}+\frac{4\dot
{\textsf{A}}\dot
{\textsf{B}}}{{\textsf{A}}^3{\textsf{B}}}-\frac{4{{\textsf{C}}'^2\textsf{A}}''}{{\textsf{A}}{\textsf{B}}^4}
\right.\right.\\\nonumber&-&\left.\left.\frac{4{\textsf{A}}'^2 \dot
{\textsf{C}}^2}{{\textsf{A}}^4{\textsf{B}}^2}-\frac{4\ddot
{\textsf{B}} \dot {\textsf{C}}^2}{{\textsf{A}}^4
{\textsf{B}}}+\frac{4\ddot
{\textsf{B}}{\textsf{C}}'^2}{{\textsf{A}}^2{\textsf{B}}^3}+\frac{4{\textsf{A}}''\dot
{\textsf{C}}^2}{{\textsf{A}}^3{\textsf{B}}^2}+\frac{8\ddot
{\textsf{C}}{\textsf{C}}''}{{\textsf{A}}^2{\textsf{B}}^2}-\frac{4\dot
{\textsf{B}}^2{\textsf{C}}'^2}{{\textsf{A}}^2{\textsf{B}}^4}\right)f_{G}\right.\\\nonumber
&+&\left.\left(\frac{4\ddot {\textsf{B}}{\textsf{C}}\dot
{\textsf{C}}}{{\textsf{A}}^4{\textsf{B}}}-\frac{12\dot
{\textsf{A}}\dot {\textsf{B}} \dot
{\textsf{C}}{\textsf{C}}}{{\textsf{A}}^5{\textsf{B}}}-\frac{4\dot
{\textsf{A}}{\textsf{B}}'{\textsf{C}}
{\textsf{C}}'}{{\textsf{A}}^3{\textsf{B}}^3}+\frac{8{\textsf{A}}'{\textsf{C}}'\dot
{\textsf{B}}}{{\textsf{A}}^3{\textsf{B}}^3}+\frac{4{\textsf{A}}'{\textsf{B}}'\dot
{\textsf{C}}{\textsf{C}}}{{\textsf{A}}^3{\textsf{B}}^3}
-\frac{4}{{\textsf{A}}^3}\right.\right.\\\nonumber
&\times&\left.\left.\frac{{\textsf{A}}''\dot
{\textsf{C}}{\textsf{C}}}{{\textsf{B}}^2}+\frac{4\dot
{\textsf{B}}\ddot {\textsf{C}}{\textsf{C}}
}{{\textsf{A}}^4{\textsf{B}}}-\frac{12{\textsf{A}}'{\textsf{C}}\dot
{\textsf{C}}'}{{\textsf{A}}^3{\textsf{B}}^2}+\frac{4\dot
{\textsf{A}}{\textsf{C}}{\textsf{C}}''}{{\textsf{A}}^3{\textsf{B}}^2}
+\frac{12{\textsf{C}}\dot
{\textsf{C}}{\textsf{A}}'^2}{{\textsf{A}}^4{\textsf{B}}^2}\right)\dot
f_{G}+\left(8\right.\right.\\\nonumber
&\times&\left.\left.\frac{{\textsf{A}}'\dot {\textsf{C}}\dot
{\textsf{B}}{\textsf{C}}}{{\textsf{A}}^3{\textsf{B}}^3}
+\frac{4{\textsf{B}}'{\textsf{C}}\ddot
{\textsf{C}}}{{\textsf{B}}^3{\textsf{A}}^2}+\frac{4{\textsf{C}}\dot
{\textsf{A}}{\textsf{C}}'\dot
{\textsf{B}}}{{\textsf{A}}^3{\textsf{B}}^3}
-\frac{12{{\textsf{C}}'{\textsf{C}}\textsf{A}}'{\textsf{B}}'}{{\textsf{A}}
{\textsf{B}}^5}-\frac{4{\dot
{\textsf{C}}{\textsf{C}}\textsf{B}}'\dot
{\textsf{A}}}{{\textsf{B}}^3{\textsf{A}}^3}+\frac{4}{{\textsf{B}}^4}\right.\right.\\\nonumber
&\times&\left.\left.
\frac{{{\textsf{C}}{\textsf{C}}'\textsf{A}}''}{{\textsf{A}}}+\frac{8\dot
{\textsf{B}}{\textsf{C}}'\dot
{\textsf{C}}'}{{\textsf{A}}^2{\textsf{B}}^3}-\frac{4\ddot
{\textsf{B}}{\textsf{C}}'{\textsf{C}}}{{\textsf{B}}^3{\textsf{A}}^2}-\frac{12\dot
{\textsf{B}}\dot
{\textsf{C}}'{\textsf{C}}}{{\textsf{B}}^3{\textsf{A}}^2}
+\frac{4{{\textsf{C}}\textsf{A}}'{\textsf{C}}''}{{\textsf{B}}^4{\textsf{A}}}
+\frac{12{\textsf{A}}'^2}{{\textsf{A}}^2}\right.\right.\\\nonumber
&\times&\left.\left.\frac{{\textsf{C}}\dot
{\textsf{C}}}{{\textsf{B}}^4}\right)f'_{G}+\left(\frac{4\dot
{\textsf{A}}\dot
{\textsf{C}}{\textsf{C}}}{{\textsf{A}}^3{\textsf{B}}^2}
+\frac{4{\textsf{A}}'{\textsf{C}}'{\textsf{C}}}{{\textsf{A}}{\textsf{B}}^4}-\frac{4\ddot
{\textsf{C}}{\textsf{C}}}{{\textsf{A}}^2{\textsf{B}}^2}\right)f''_{G}
+\left(\frac{12{\textsf{C}}\dot
{\textsf{C}}'}{{\textsf{A}}^2{\textsf{B}}^2}
-\frac{12{\textsf{C}}}{{\textsf{A}}^2}\right.\right.\\\label{100c}
&\times&\left.\left.\frac{\dot
{\textsf{B}}{\textsf{C}}'}{{\textsf{B}}^3}-\frac{12{\textsf{A}}'{\textsf{C}}\dot
{\textsf{C}}}{{\textsf{A}}^3{\textsf{B}}^2}\right)\dot
f'_{G}+\left(\frac{4\dot{\textsf{B}}\dot
{\textsf{C}}{\textsf{C}}}{{\textsf{A}}^4{\textsf{B}}}
+\frac{4{\textsf{B}}'{\textsf{C}}'{\textsf{C}}}{{\textsf{A}}^2{\textsf{B}}^3}
-\frac{4{\textsf{C}}{\textsf{C}}''}{{\textsf{A}}^2{\textsf{B}}^2}\right)\ddot
f_{G}\right].
\end{eqnarray}
The modified expressions of $\texttt{Z}_1$ and $\texttt{Z}_2$ are
given as
\begin{align}\nonumber
\texttt{Z}_{1}&=\frac{f_{T}}{8\pi-f_{T}}\bigg[\big(\ln f_T
\big)^{.}\bigg(\frac{{\mathfrak{u}}}{{\textsf{A}}}+\frac{\mathbf{s}^2}{8\pi
\textsf{C}^4\textsf{A}}+
\frac{T^{00\textsf{(GT)}}}{{\textsf{A}}^3}+\frac{{\mathfrak{p}}}{{\textsf{A}}}\bigg)+2\bigg(\frac{{\zeta}}{{\textsf{B}}}
-\frac{T^{01\textsf{(GT)}}}{{\textsf{A}}{\textsf{B}}^2}\bigg)\\\nonumber
&+\frac{1}{2{\textsf{A}}}\bigg({\mathfrak{u}}+3{\mathfrak{p}}\bigg)^.+(\ln
f_T)'\bigg(\frac{{\zeta}}{{\textsf{B}}}-\frac{T^{01\textsf{(GT)}}}{{\textsf{A}}{\textsf{B}}^2}\bigg)
+\bigg(\frac{2{\mathfrak{u}}}{{\textsf{A}}}+\frac{\mathbf{s}^2}{4\pi
\textsf{C}^4\textsf{A}}+\frac{2T^{00\textsf{(GT)}}}{{\textsf{A}}^3}
\\\label{100d}&+\frac{{\mathfrak{p}}}{{\textsf{A}}}\bigg)^.\bigg],\\\nonumber
\texttt{Z}_{2}&=\frac{f_{T}}{8\pi-f_{T}}\bigg[\big(\ln f_T
\big)'\bigg(\frac{-{\mathfrak{p}}_r}{{\textsf{B}}}-\frac{\mathbf{s}^2}{8\pi
\textsf{C}^4\textsf{B}}+\frac{T^{11\textsf{(GT)}}}{{\textsf{B}}^3}
+\frac{{\mathfrak{p}}}{{\textsf{B}}}\bigg)+2\bigg(\frac{T^{01\textsf{(GT)}}}{{\textsf{A}}^2{\textsf{B}}}-\frac{{\zeta}}{{\textsf{A}}}\bigg)\\\nonumber
&-\frac{1}{2}\frac{\bigg(3{\mathfrak{p}}+{\mathfrak{u}}\bigg)'}{{\textsf{B}}}+\bigg(\frac{-2{\mathfrak{p}}_r}{{\textsf{B}}}-\frac{\mathbf{s}^2}{4\pi
\textsf{C}^4\textsf{B}} -\frac{2T^{11\textsf{(GT)}}}{{\textsf{B}}^3}
+\frac{{\mathfrak{p}}}{{\textsf{B}}}\bigg)'+\bigg(\frac{-{\zeta}}{{\textsf{A}}}
+\frac{T^{01\textsf{(GT)}}}{{\textsf{A}}^2{\textsf{B}}}\bigg)\\\label{100e}&\times(\ln
f_T)^.\bigg].
\end{align}

\section*{Appendix B}
\renewcommand{\theequation}{B\arabic{equation}}
\setcounter{equation}{0} In scalar functions, the extra curvature
terms are described as follows
\begin{eqnarray}\nonumber
\textsf{M}^{\textsf{(GT)}}&=&
2\left[R_{\mu\tau}R^{\mu}_{\vartheta}f_{G}+
R^{\mu\nu}R_{\mu\tau\nu\vartheta}f_{G}-\frac{1}{2}RR_{\vartheta\tau}f_{G}-\frac{1}{2}R_{\tau\mu\nu
n}R^{\mu\nu
n}_{\vartheta}f_{G}\right.\\\nonumber&+&\left.\frac{1}{2}R
\nabla_{\vartheta}\nabla_{\tau}f_{G}+R_{\vartheta\tau}\Box
f_{G}-R^{\mu}_{\vartheta}\nabla_{\tau}\nabla_{\mu}f_{G}-R^{\mu}_{\tau}\nabla_{\vartheta}\nabla_{\mu}f_{G}\right.\\\nonumber
&-&\left.R_{\mu\tau
\nu\vartheta}\nabla^{\mu}\nabla^{\nu}f_{G}\right]g^{\vartheta\tau}+2\left[-R_{\mu\delta}R^{\mu}_{\vartheta}f_{G}-
R^{\mu\nu}R_{\mu\delta
\nu\vartheta}f_{G}+\frac{1}{2}RR_{\vartheta\delta}f_{G}\right.\\\nonumber&+&\left.\frac{1}{2}R_{\delta
\mu\nu n}R^{\mu\nu n}_{\vartheta}f_{G}-R_{\delta\vartheta}\Box
f_{G}+R_{\mu\delta\nu\vartheta}\nabla^{\nu}\nabla^{\mu}f_{G}+R^{\mu}_{\vartheta}\nabla_{\mu}\nabla_{\delta}f_{G}
\right.\\\nonumber&+&\left.R^{\mu}_{\delta}\nabla_{\vartheta}\nabla_{\mu}
f_{G}-\frac{1}{2}R
\nabla_{\vartheta}\nabla_{\delta}f_{G}\right]v_{\tau}v^{\delta}g^{\vartheta\tau}+2\left[-R_{\mu\tau}R^{\mu\gamma}f_{G}\right.\\\nonumber&+&\left.
\frac{1}{2}R_{\tau\mu\nu n}R^{\mu\nu
n\gamma}f_{G}-+\frac{1}{2}RR^{\gamma}_{\tau}f_{G}
R^{\mu\nu}R^{\gamma}_{\mu\tau
\nu}f_{G}+R^{\mu\gamma}\nabla_{\tau}\nabla_{\mu}f_{G}\right.\\\nonumber&-&\left.\frac{1}{2}R
\nabla^{\gamma}\nabla_{\tau}f_{G}+R^{\gamma}_{\mu\tau
\nu}\nabla^{\nu}\nabla^{\mu}f_{G}-R^{\gamma}_{\tau}\Box
f_{G}+R^{\mu}_{\tau}\nabla_{\mu}\nabla^{\gamma}f_{G}\right]v_{\vartheta}v_{\gamma}g^{\vartheta\tau}\\\nonumber
&+&2\left[R^{\mu\gamma}R_{\mu\delta}f_{G}+
R^{\mu\nu}R^{\gamma}_{\mu\delta \nu}f_{G}+\frac{1}{2}R
\nabla_{\delta}\nabla^{\gamma}f_{G}-\frac{1}{2}R^{\mu\nu n
\gamma}R_{\delta \mu\nu
n}f_{G}\right.\\\nonumber&+&\left.R^{\gamma}_{\delta}\Box f_{G}
-R^{\mu}_{\delta}\nabla_{\mu}\nabla^{\gamma}f_{G}-R^{\mu\gamma}\nabla_{\mu}\nabla_{\delta}f_{G}-R^{\gamma}_{\mu\delta
\nu}\nabla^{\nu}\nabla^{\mu}f_{G}-\frac{1}{2}RR^{\gamma}_{\delta}f_{G}\right]
\\\nonumber &\times&g_{\vartheta\tau}v_{\gamma}v^{\delta}g^{\vartheta\tau}-\left[4R_{\mu m}R^{\mu m}f_{G}
-2R^{2}f_{G}+4R^{m}_{\mu m
\nu}R^{\mu\nu}f_{G}\right.\\\nonumber&-&\left.2R^{l}_{\mu \nu
n}R^{\mu \nu
n}_{l}f_{G}+16R^{\mu\nu}\nabla_{\nu}\nabla_{\mu}f_{G}-4R\Box f_{G}
-4R^{\mu
m}\nabla_{\mu}\nabla_{m}f_{G}\right.\\\nonumber&-&\left.4R^{m}_{\mu
m \nu}\nabla^{\nu}\nabla^{\mu}f_{G}-4R^{\nu
l}\nabla_{\nu}\nabla_{l}f_{G}\right]-6R\Box f_{G}-
\frac{1}{2}f+12R^{\mu\nu}\nabla_{\mu}\nabla_{\nu}f_{G},
\end{eqnarray}
\begin{eqnarray}\nonumber
\textsf{J}^{\textsf{(GT)}}_{(\vartheta\tau)} &=& \left[2R_{\mu
d}R^{\mu}_{c}f_{G}+2R^{\mu\nu}R_{\mu d \nu
c}f_{G}-RR_{cd}f_{G}-R_{d\mu\nu n}R^{\mu\nu n}_{c}f_{G}+2R_{cd}\Box
f_{G}\right.\\\nonumber&+&\left.R\nabla_{d}\nabla_{c}f_{G}-2R^{\mu}_{d}\nabla_{\mu}\nabla_{c}f_{G}-2R^{\mu}_{c}\nabla_{\mu}\nabla_{d}f_{G}
-2R_{\mu d\nu c}\nabla^{\mu}\nabla^{\nu}f_{G}\right]
h^{c}_{\vartheta}h^{d}_{\tau}\\\nonumber&+&2\left[R^{\mu\gamma}R_{\mu\delta}f_{G}+
R^{\gamma}_{\mu\delta
\nu}R^{\mu\nu}f_{G}-\frac{1}{2}RR^{\gamma}_{\delta}f_{G}-\frac{1}{2}R_{\delta
\mu\nu n}R^{\mu\nu n
\gamma}f_{G}\right.\\\nonumber&+&\left.R^{\gamma}_{\delta}\Box
f_{G}-R^{\mu\gamma}\nabla_{\mu}\nabla_{\delta}f_{G}+\frac{1}{2}R
\nabla_{\delta}\nabla^{\gamma}f_{G}
-R^{\mu}_{\delta}\nabla^{\gamma}\nabla_{\mu}f_{G}\right.\\\nonumber&-&\left.R^{\gamma}_{\mu\delta
\nu}\nabla^{\mu}\nabla^{\nu}f_{G}\right]h_{\vartheta\tau}v_{\gamma}v^{\delta}-2\left[R_{\mu\tau}R^{\mu}_{\vartheta}f_{G}+
R^{\mu\nu}R_{\mu\tau
\nu\vartheta}f_{G}-\frac{1}{2}RR_{\vartheta\tau}f_{G}\right.\\\nonumber&-&\left.\frac{1}{2}R_{\tau
\mu\nu n}R^{\mu\nu n}_{\vartheta}f_{G}+R_{\vartheta\tau}\Box
f_{G}+\frac{1}{2}R
\nabla_{\vartheta}\nabla_{\tau}f_{G}-R^{\mu}_{\vartheta}\nabla_{\tau}\nabla_{\mu}f_{G}\right.\\\nonumber&-&\left.
R^{\mu}_{\tau}\nabla_{\vartheta}\nabla_{\mu}f_{G}-R_{\mu\tau
\nu\vartheta}\nabla^{\mu}\nabla^{\nu}f_{G}\right]-2\left[-R_{\mu\delta}R^{\mu}_{\vartheta}f_{G}-
R^{\mu\nu}R_{\mu\delta
\nu\vartheta}f_{G}\right.\\\nonumber&+&\left.\frac{1}{2}RR_{\vartheta\delta}f_{G}+\frac{1}{2}R_{\delta
\mu\nu n}R^{\mu\nu n}_{\vartheta}f_{G}-R_{\delta\vartheta}\Box
f_{G}+R_{\mu\delta
\nu\vartheta}\nabla^{\mu}\nabla^{\nu}f_{G}\right.\\\nonumber&-&\left.\frac{1}{2}R
\nabla_{\vartheta}\nabla_{\delta}f_{G}+R^{\mu}_{\vartheta}\nabla_{\delta}\nabla_{\mu}f_{G}
+R^{\mu}_{\delta}\nabla_{\vartheta}\nabla_{\mu}f_{G}\right]v_{\tau}v^{\delta}-2\left[-R_{\mu\tau}R^{\mu\gamma}f_{G}\right.
\\\nonumber&-&\left.R^{\mu\nu}R^{\gamma}_{\mu\tau
\nu}f_{G}+\frac{1}{2}RR^{\gamma}_{\tau}f_{G}+\frac{1}{2}R_{\tau
\mu\nu n}R^{\mu\nu n\gamma}f_{G}-R^{\gamma}_{\tau}\Box
f_{G}\right.\\\nonumber&+&\left.R^{\mu\gamma}\nabla_{\mu}\nabla_{\tau}f_{G}+R^{\mu}_{\tau}\nabla_{\mu}\nabla^{\gamma}f_{G}-\frac{1}{2}R
\nabla^{\gamma}\nabla_{\tau}f_{G}+R^{\gamma}_{\mu\tau
\nu}\nabla^{\mu}\nabla^{\nu}f_{G}\right]v_{\vartheta}v_{\gamma}\\\nonumber&-&2\left[
R^{\gamma}_{\mu\delta
\nu}R^{\mu\nu}f_{G}+R_{\mu\delta}R^{\mu\gamma}f_{G}-\frac{1}{2}RR^{\gamma}_{\delta}f_{G}-\frac{1}{2}R^{\mu\nu
n \gamma}R_{\delta \mu\nu
n}f_{G}\right.\\\nonumber&+&\left.R^{\gamma}_{\delta}\Box
f_{G}+\frac{1}{2}R
\nabla_{\delta}\nabla^{\gamma}f_{G}-R^{\mu}_{\delta}\nabla_{\mu}\nabla^{\gamma}f_{G}-R^{\mu\gamma}\nabla_{\mu}\nabla_{\delta}f_{G}
\right.\\\nonumber&-&\left.R^{\gamma}_{\mu\delta
\nu}\nabla^{\nu}\nabla^{\mu}f_{G}\right]v_{\gamma}v^{\delta}g_{\vartheta\tau},
\end{eqnarray}
\begin{eqnarray}\nonumber
\textsf{Q}^{\textsf{(GT)}} &=&\left[\frac{1}{2}R_{\mu\epsilon}R^{\mu
p}f_{G}+\frac{1}{2}R^{\mu\nu}R^{p}_{\mu\epsilon
\nu}f_{G}-\frac{1}{4}RR^{p}_{\epsilon}f_{G}-\frac{1}{4}R_{\epsilon
\mu\nu n}R^{\mu\nu n p}f_{G}\right.\\\nonumber&+&\left.\frac{1}{2}
R^{p}_{\epsilon}\Box
f_{G}+\frac{1}{4}R\nabla^{p}\nabla_{\epsilon}f_{G}-\frac{1}{4}R^{\mu
p}\nabla_{\epsilon}\nabla_{\mu}f_{G}-\frac{1}{2}R^{\mu}_{\epsilon}\nabla^{p}\nabla_{\mu}
f_{G}\right.\\\nonumber&-&\left.\frac{1}{2}R^{p}_{\mu\epsilon
\nu}\nabla^{\mu}\nabla^{\nu}f_{G}\right]g^{\vartheta\tau}\epsilon_{p
\delta\tau}\epsilon^{\epsilon\delta}_{\vartheta}+\left[-\frac{1}{2}R_{\mu\delta}R^{\mu
p}f_{G}-\frac{1}{2}R^{\mu\nu}R^{p}_{\mu\delta \nu}
f_{G}\right.\\\nonumber&+&\left.\frac{1}{4}R^{p}_{\delta}Rf_{G}-\frac{1}{2}R^{p}_{\delta}\Box
f_{G}+\frac{1}{4}R^{\mu\nu np}R_{\delta \mu\nu
n}f_{G}-\frac{1}{4}R\nabla^{p}\nabla_{\delta}f_{G}\right.\\\nonumber&+&\left.\frac{1}{2}R^{\mu}_{\delta}\nabla^{p}\nabla_{\mu}f_{G}+\frac{1}{4}R^{\mu
p}\nabla_{\delta}\nabla_{\mu}f_{G} +\frac{1}{2}R^{p}_{\mu\delta
\nu}\nabla^{\mu}\nabla^{\nu}f_{G}\right]g^{\vartheta\tau}\epsilon_{p\epsilon
\tau}\epsilon^{\epsilon\delta}_{\vartheta}\\\nonumber&+&\left[-\frac{1}{2}R_{\mu\epsilon}R^{\mu\gamma}f_{G}-\frac{1}{2}R^{\mu\nu}R^{\gamma}_{\mu\epsilon
\nu}f_{G}
+\frac{1}{4}RR^{\gamma}_{\epsilon}f_{G}+\frac{1}{4}R_{\epsilon
\mu\nu n}R^{\mu\nu
n\gamma}f_{G}\right.\\\nonumber&-&\left.\frac{1}{2}R^{\gamma}_{\epsilon}\Box
f_{G}+\frac{1}{4}R^{\mu\gamma}\nabla_{\epsilon}\nabla_{\mu}f_{G}-\frac{1}{4}R\nabla^{\gamma}\nabla_{\epsilon}f_{G}
+\frac{1}{2}R^{\mu}_{\epsilon}\nabla_{\mu}\nabla^{\gamma}
f_{G}\right.\\\nonumber&+&\left.\frac{1}{2}R^{\gamma}_{\mu\epsilon
\nu}\nabla^{\nu}\nabla^{\mu}
f_{G}\right]g^{\vartheta\tau}\epsilon_{\delta\gamma\tau}\epsilon^{\epsilon\delta}_{\vartheta}+\left[\frac{1}{2}R^{\mu\nu}R^{\gamma}_{\mu\delta
\nu}f_{G}+\frac{1}{2}R^{\mu\gamma}R_{\mu\delta}f_{G}
\right.\\\nonumber&-&\left.\frac{1}{4}RR^{\gamma}_{\delta}f_{G}+\frac{1}{2}R^{\gamma}_{\delta}\Box
f_{G}-\frac{1}{4}R^{\mu\nu n\gamma}R_{\delta \mu\nu
n}f_{G}+\frac{1}{4}R\nabla_{\delta}\nabla^{\gamma}f_{G}\right.\\\nonumber&-&\left.\frac{1}{4}R^{\mu\gamma}\nabla_{\delta}\nabla_{\mu}f_{G}
-\frac{1}{2}R^{\gamma}_{\mu\delta
\nu}\nabla^{\mu}\nabla^{\nu}f_{G}-\frac{1}{2}R^{\mu}_{\delta}\nabla^{\gamma}\nabla_{\mu}f_{G}\right]
g^{\vartheta\tau}\epsilon_{\epsilon\gamma\tau}\epsilon^{\epsilon\delta}_{\vartheta}\\\nonumber&-&
12R^{\mu\nu}\nabla_{\nu}\nabla_{\mu}f_{G} +6R\Box
f_{G}+\left[4R_{\mu\nu}R^{\mu\nu}f_{G}+\left(\mu+{\mathfrak{p}}\right)f_{T}\right.\\\nonumber&+&\left.4R^{\mu\nu}R^{m}_{\mu\nu
m}f_{G}-2R^{l}_{\mu\nu n}R^{\mu\nu n}_{l}f_{G}-2R^{2}f_{G}-4R\Box
f_{G}\right.\\\nonumber&+&\left.16R^{\mu\nu}\nabla_{\nu}\nabla_{\mu}f_{G}-4R^{\mu
l}\nabla_{l}\nabla_{\mu}f_{G}-4R^{\mu m}\nabla_{\mu}\nabla_{m}f_{G}
\right.\\\nonumber&-&\left.4R^{m}_{\mu m
\nu}\nabla^{\nu}\nabla^{\mu}f_{G}\right]+\frac{f}{2}.
 \end{eqnarray}
\\
\textbf{Data Availability Statement:} This manuscript has no
associated data.

\end{document}